\documentclass[]{aa}
\usepackage{natbib}
\bibpunct{(}{)}{;}{a}{}{,}  
\defcitealias{shakura1973black}{SS73} 
\defcitealias{illenseer2015self}{ID15} 
\usepackage{graphicx}
\usepackage{txfonts}
\usepackage{comment}
\usepackage[]{hyperref}
\begin{document}

   \title{Accretion disk dynamics}

   \subtitle{$\alpha$-viscosity in self-similar self-gravitating models}

 \author{Marcus Kubsch
          \inst{1}
          \and
          Tobias F. Illenseer\inst{1}\
        \and
        Wolfgang J. Duschl \inst{1,} \inst{2}
          }

   \institute{ Institut f\"ur Theoretische Physik und Astrophysik, Chrisitan-Albrechts-Universit\"at zu Kiel,
              Leibnizstra\ss e 15, 24118 Kiel, Germany\\
              \email{\\ mmeissner@astrophysik.uni-kiel.de, tillense@astrophysik.uni-kiel.de, wjd@astrophysik.uni-kiel.de}
         \and
             Steward Observatory, The University of Arizona,
              933 N. Cherry Ave., Tucson, AZ 85721, USA
             }

   \date{Received September 15, 1996; accepted March 16, 1997}

 
  \abstract
   {}
   {We investigate  the suitability of $\alpha$-viscosity  in self-similar models for self-gravitating disks with a focus on active galactic nuclei (AGN)  disks.}
   {We use a self-similar approach to simplify the partial differential equations arising from the evolution equation, which are then solved using numerical standard procedures.}
   {We find a self-similar solution for the dynamical evolution of self-gravitating $\alpha$-disks and derive the significant quantities. In the Keplerian part of the disk our model is consistent with standard stationary $\alpha$-disk theory, and self-consistent throughout the self-gravitating regime. Positive accretion rates throughout the disk demand a high degree of self-gravitation. Combined with the temporal decline of the accretion rate and its low amount, the model prohibits the growth of large central masses. }
   {$\alpha$-viscosity cannot account for the evolution of the whole mass spectrum of super-massive black holes (SMBH) in AGN. However, considering the involved scales it seems suitable for modelling protoplanetary disks.}

   \keywords{accretion, accretion disks -- turbulence -- hydrodynamics -- methods: analytical
               }

   \maketitle

\section{Introduction}
Over half a century ago, the rise of radio astronomy lead to the discovery of quasars. The extreme luminosity of those point like sources baffled astronomers at first. At the time, nuclear fusion, the well-known energy source of stars, was deemed \emph{the} energy source of the universe. However, it quickly became apparent that nuclear fusion could not account for the tremendous amounts of energy released. The solution was proposed by \cite{zeldovich1964fate}, \cite{salpeter1964accretion}, and \cite{bell}: the release of gravitational energy from material falling from a rotating disk onto a massive central object at its center, i.e. an accretion disk. Since then, accretion disks have become the paradigm and a major topic of astrophysical research. The huge interest in these disks is fueled by the fact that they can explain phenomena ranging from the evolution of super-massive black holes (SMBH) in active galactic nuclei (AGN) to star- and planet formation. 

However, owing to the law of momentum conservation, a successful accretion process depends on a mechanism that  directs angular momentum from the inner regions to the outer regions. The mechanism of choice is viscosity. Originally derived by \cite{goldreich1967differential} for rotating stars\footnote{See e.g. \cite{kato2008black} for an application to accretion disks.}, it is now generally agreed  that molecular viscosity, because its  corresponding viscous timescale is too long to accord for the observed quasar luminosities, is a negligible process. \cite{bell} himself considered magnetic fields resulting from non-uniform rotation to be the best  candidates for angular momentum transport, which \cite{mri} confirmed, although under the condition that charged particles and a small magnetic field acting as a seed exist in the accretion disk. However, if  the velocity field in an accretion disk and the corresponding timescales are considered,  extremely high Reynolds numbers are obtained \citep[e.g.][]{lynden1974evolution}. In consequence, turbulence is considered to emerge. Turbulent flows and the resulting eddies offer a transport mechanism for angular momentum and mass (\cite{weizsacker1948rotation} \& \cite{lust1952entwicklung}) which is independent of the existence of magnetic fields.

Consequently,  some kind of viscosity prescription $\nu$ is needed to describe accretion disks. The widely used $\alpha$ ansatz by \cite{shakura1973black} (hereafter SS73) produces physically unreasonable results such as completely isothermal disks when applied to geometrically thin and stationary AGN disks \citep{Duschl:433740}. Furthermore, \cite{alpha_ineffective} found that the $\alpha$ ansatz is not efficient enough to fuel the observed rapid mass growth of AGN \citep{fan2003survey}. Thus, \cite{illenseer2015self} (hereafter ID15) neglected the $\alpha$ ansatz and used - among others - the more general and physically faithful $\beta$ description \citep{Duschl:433740} when they developed their dynamical model for self-gravitating accretion disks. 

However, since the viscosity prescription is left undetermined in the most general form of the self-similar model by \citetalias{illenseer2015self}, we  investigate  the consequences of applying the $\alpha$ prescription in this paper. Accordingly, we  derive a disk equation using $\alpha$-viscosity and apply the methodology of similarity solutions to arrive at a system of ordinary differential equations (ODEs) which describe the time evolution of the disk. These equations  are solved using numerical standard procedures. In turn, these results are discussed, which will finally  allow us to review the suitability of using the $\alpha$ ansatz to model AGN. 

It is out of the scope of this paper to provide a detailed discussion of the numerical schemes and the concept of utilizing self-similarity to solve differential equations. For a short overview of the latter see \cite{dresner1998applications}; for a comprehensive treatment we recommend \cite{bluman2010applications}. 
 
\section{ Disk evolution equation}
For a thin, axisymmetric disk which is in hydrostatic balance in the vertical direction, \citetalias{illenseer2015self} derived the following disk evolution equation
\begin{align} \label{original}
-r^4\Omega^2\frac{\partial\Omega}{\partial t}=\frac{\partial}{\partial r}\left (\nu r^3 \Omega^3 x(2x+3) \right )
\end{align} 
with angular velocity $\Omega=\frac{v_{\varphi}}{r}$ and kinematic viscosity $\nu(r,t)$. In combination with 
\begin{align}
\label{x}
x=\frac{r}{\Omega} \frac{\partial \Omega}{\partial r} =\frac{\partial  \ln  \Omega}{\partial  \ln  r},
\end{align}
which gives the local power law exponent of the rotation law, this second-order non-linear partial differential equation (PDE) characterizes the transport processes of angular velocity under the influence of self-gravity and viscous friction. With given boundary and initial conditions and a viscosity prescription   this equation can be solved. 
\subsection{Viscosity prescription} 
Viscosity is important for the transport of angular momentum and the heating of the disk. A widely used prescription is achieved\footnote{This is just a variant of the more common form $\nu = \alpha c_s H$, where $H$ is the scale height of the disk.} via the $\alpha$ parametrization of \citetalias{shakura1973black}\footnote{For a modern treatment of the standard $\alpha$ theory, see e.g. \cite{kato2008black} or \cite{frank2002accretion}.},
\begin{align}
T_{r \varphi}& = -\alpha \Pi  \label{alt_alpha} ,
\end{align} 
which is related to the disk model via the $r \varphi$ component of the sheer stress tensor \citepalias{shakura1973black},
\begin{align}
T_{r \varphi}&=\nu \Sigma r \frac{d\Omega}{dr}  \label{2}.
\end{align} 
Here, $\Sigma$ and $\Pi$  are the respective vertically integrated quantities of  density  $\rho$ and pressure $p$ which -- assuming vertical isothermy -- are connected via
\begin{align}
c_s=\sqrt{\frac{\gamma p}{\rho}} = \sqrt{\frac{\gamma \Re T}{m_{\text{mol}}}} \label{cs_t}=\sqrt{\frac{\Pi}{\Sigma}},
\end{align} 
where $c_s$ denotes the speed of sound; $\Re$ the gas constant; $m_{\text{mol}}$ molar mass; $\gamma$ the adiabatic coefficient, which can be assumed to be of the order of 1;  and $T$ the temperature\footnote{$T$, $\rho$, and $p$ denote the corresponding quantities in the equatorial plane.}. 
Because  $ c_s\propto\sqrt{T} $, the heating and cooling mechanisms of the disk will eventually be of interest.

By equating Eqs. \eqref{alt_alpha} and \eqref{2}, solving for $\nu$, and using Eq. \eqref{cs_t} with vertically integrated quantities, we obtain
\begin{align}
\nu=-\frac{\alpha c_{s}^{2}}{ r \,\partial_r \Omega}, \label{nu}
\end{align} 
which can now be inserted into Eq. \eqref{original}: 
\begin{align} \label{4}
-r^4\Omega^2\frac{\partial \Omega}{\partial t}=-\frac{\partial}{\partial r}\left (\frac{\alpha c_{s}^{2}}{ r \,\partial_r \Omega} r^3 \Omega^3 x(2x+3) \right ).
\end{align}
Applying the definition of $x$ (Eq. \eqref{x}) leads to the  expression 
\begin{align}
r^4\Omega^2 \frac{\partial \Omega}{\partial t}=\alpha \frac{\partial}{\partial r}\left (c_s^2 r^3 \Omega^2(2x+3) \right ).
\end{align} 

Since the speed of sound is dependent on the radius, a relationship between $c_s$ and the other parameters of the system is constructed to eliminate $c_s$ from the evolution equation. This can be achieved via the energy equation. Presuming a vertically optical thin disk, the effective temperature can be assumed to be equal to the central temperature ($T=T_{\text{eff}}$), i.e. generated heat is immediately radiated locally and consequently the thermal timescale must be much shorter than the matter diffusion timescale \citep{shakura1976theory}. Thus the viscous dissipation rate \citepalias{shakura1973black} can be equated to the effective temperature of a black-body spectrum and -- using the vertically integrated form of Eq. \eqref{cs_t} -- can be expressed in terms of $\Sigma$ and the speed of sound:  
\begin{align}
2\sigma T^4&=-\alpha \Pi r \frac{\partial \Omega}{\partial r}=-\alpha c_s^2 \Sigma r \frac{\partial \Omega}{\partial r} \label{t_eff}.
\end{align}   
With help of  the relation
\begin{align}
\Sigma=\frac{r \Omega^2(2x+3)}{2\pi G} \label{sig}
\end{align}
derived in \citetalias{illenseer2015self} and Eq. \eqref{cs_t}   $\Sigma$  and $T$ can be eliminated from Eq. \eqref{t_eff}, which gives   
\begin{align}
c_s^2=-\left( \frac{\alpha \gamma \Re^4}{4 \sigma \pi G m_{\text{mol}}^4} \left ( r^2 \Omega^2 (2x+3) \partial_r \Omega \right )\right )^\frac{1}{3}. \label{cs}
\end{align}  
This result can now be used to eliminate the speed of sound from the evolution equation, which gives
\begin{align}
r^4\Omega^2\frac{\partial \Omega}{\partial t}&=\alpha \frac{\partial}{\partial r}\left (-\left( \frac{\alpha \gamma \Re^4}{4 \sigma \pi G m_{\text{mol}}^4} \left ( \partial_r \Omega \right )\right )^\frac{1}{3}r^{\frac{11}{3}} \Omega^{\frac{8}{3}} (2x+3)^{\frac{4}{3}} \right ).
\end{align} 
After a short calculation using the definition of  $x$ and setting
\begin{align}
\eta=\left (\frac{\alpha^4 \gamma \Re^4}{4 \sigma \pi G m_{\text{mol}}^4} \right )^{\frac{1}{3}}, \label{gamma}
\end{align} 
 the subsequent result is obtained: 
\begin{align} \label{6}
r^4 \Omega^2 \frac{\partial \Omega}{\partial t}=-\eta \frac{\partial}{\partial r}\left ( r^{\frac{10}{3}} \Omega^3 x^{\frac{1}{3}} (2x+3)^{\frac{4}{3}} \right ). 
\end{align} 

To remove unnecessary complexities in the further derivations, dimensionless scales for length, mass, and time ($\tilde r$, $\tilde M$, and $\tilde t$) are be used in the next sections. The dimensions in SI units of the basic quantities involved are as follows: 
\begin{align}
[\Omega]=\frac{1}{\text{s}}, \;
[t]=\text{s}, \;
[r]=\text{m},\;
[\eta]=\frac{\text{m}^{\frac{5}{3}}}{\text{s}}.
\end{align} 
Now, a scaling relation for  $\tilde t$ corresponding to 
\begin{align}
\hat t = \sqrt{\frac{\hat r^3}{G \hat M} } \label{dimless} 
\end{align}
is defined which  allows  the dimensionless solutions to be rescaled. Except for equations containing Newton's constant, which has to be set to unity, all expressions preserve their form\footnote{To provide an easily legible type, the dimensionless equations in the next sections are written without tildes.}. When the dimensionless problem has been solved, it is possible to return to physical quantities via the scaling transformations arising from Eq. \eqref{dimless}. Two out of the three scales (mass, length, time) are sufficient to be able to calculate the third. 
Consequently, $\eta$ becomes a dimensionless parameter $\tilde \eta$. Its value depends on the dimensions of the actual problem via 
\begin{align}
\tilde \eta = \eta \frac{\hat t}{\hat r^{\frac{5}{3}}}.
\end{align} 
Furthermore, substituting $\tau=5\tilde \eta \tilde t$ allows  Eq. \eqref{6} to be rewritten, which yields the following partial differential equation:
\begin{align}
5 \tilde r^4 \tilde \Omega^2 \frac{\partial \tilde \Omega}{\partial  \tau }=- \frac{\partial}{\partial \tilde r}\left ( \tilde r^{\frac{10}{3}} \tilde \Omega^3 x^{\frac{1}{3}} (2x+3)^{\frac{4}{3}} \right ). 
\end{align} 
Thus, $\tilde \eta$ is basically the viscous coupling parameter equivalent to the $\beta$ used by \citetalias{illenseer2015self}. 

\subsection{Self-similar solution}
A first useful step to reduce the complexity of solving the disk evolution PDE (Eq. \eqref{6}) is to introduce new variables and rewrite the equation in those terms. In order to eliminate the roots of the radial coordinate,  the new variable  can now be defined  
\begin{align} \label{defdelta}
\varpi &= r^{\frac{5}{3}} 
\end{align} 
and plugged  into Eq. \eqref{6}. After a short calculation 
\begin{align} 
3 \varpi^2 \Omega^2 \frac{\partial \Omega}{\partial \tau}&=- \frac{\partial}{\partial \varpi} \left (  \varpi^2  \Omega^3 x^{\frac{1}{3}} (2x+3)^{\frac{4}{3}} \right ) \label{transevo} 
\end{align} 
is obtained. Obviously,  $x$   has to be transformed (as defined in Eq. \eqref{x}) as well:
\begin{align} \label{transx}
x=\left (\frac{d \ln \varpi}{d \ln r}\right ) \left ( \frac{\partial \ln \Omega}{\partial \ln \varpi} \right) = \frac{5}{3}\frac{\partial \ln \Omega}{\partial \ln \varpi}.
\end{align} 

\subsection{Scaling transformation}
The next step in applying the similarity method is to determine the group invariants. To this end, a one-parameter scaling transformation with group parameter $\lambda$ and family parameters $a, \, b, \, c$ is used:
\begin{align}
\varpi'=\lambda^a\varpi, \; \tau'=\lambda^b \tau, \; \Omega'=\lambda^c \Omega.
\end{align}   
Inserting the primed variables in Eq. \eqref{transx} gives 
\begin{align}
x=\frac{5}{3}\frac{\partial \ln \Omega}{\partial \ln \varpi}=\frac{5}{3}\frac{\varpi}{\Omega}\frac{\partial \Omega}{\partial \varpi}=\frac{5}{3}\frac{\lambda^a}{\lambda^c}\frac{\varpi'}{\Omega'}\frac{\lambda^c}{\lambda^a}\frac{\partial \Omega'}{\partial \varpi'}=\frac{5}{3}\frac{\partial \ln \Omega'}{\partial \ln \varpi'},
\end{align} 
which allows  $x$ to remain unchanged when repeating the operation for Eq. \eqref{transevo}:
\begin{align}
\frac{3}{5} \varpi'^2 \Omega'^2 \frac{\lambda^b}{\lambda^{2a} \lambda^{3c}}\frac{\partial \Omega'}{\partial \tau'}&=- \frac{\lambda^a}{\lambda^{2a} \lambda^{3c}}\frac{\partial}{\partial \varpi'} \left (  \varpi'^2  \Omega'^3 x^{\frac{1}{3}} (2x+3)^{\frac{4}{3}} \right ). 
\end{align} 
Consequently, it follows immediately that transformations of the evolution equation \eqref{transevo} are invariant if and only if $a = b$ and $c$ remains a constant. After a short calculation using these parameters, it becomes apparent that with $\kappa = \frac{c}{a}$ the following relations hold:
\begin{align}
\frac{\varpi'}{\tau'} = \frac{\varpi}{\tau},  \; \; \; \; \; \;
\Omega'\tau'^{-\kappa} = \Omega \tau^{-\kappa}.
\end{align} 
The set of group invariants 
\begin{align} \label{invariants}
\xi = \frac{\varpi}{\tau},  \; \; \; \; \; \; y(\xi)=\psi \tau^{-\kappa},
\end{align} 
where $\psi =\varpi^2\Omega^3$ allows  the original PDE (Eq. \eqref{transevo}) to be rewritten as  
\begin{align}
\frac{\partial \psi}{\partial \tau}&=-\frac{\partial}{\partial \varpi} \left ( \psi f(x) \right ), \label{transpde}
\end{align}  
where $f(x)$ collects all terms depending on $x$ and $f'$ is its derivative
\begin{align}
 f(x)&= x^{\frac{1}{3}} (2x+3)^{\frac{4}{3}} \label{f} \\ 
\frac{df}{dx}&=\frac{(2 x+3)^{\frac{1}{3}}(10 x+3)}{3 x^{\frac{2}{3}}}  .
\end{align}  
The remaining transformation of $x$ (Eq. \eqref{transx}) yields
\begin{align}
x=\frac{5}{9} \left ( \frac{\partial  \ln  \psi}{\partial  \ln \varpi}-2 \right ) . \label{lastx}
\end{align}

Now,  a set of two first-order ordinary differential equations  can be derived by applying the group invariants. Starting with the last equation \eqref{lastx}, the first step is to convert $ \frac{\partial  \ln  \psi}{\partial   \ln \varpi}$ to $ \frac{\partial  \ln  y}{\partial   \ln \xi}$. After two subsequent transformations, it is found that  $ \frac{\partial  \ln  \psi}{\partial   \ln \varpi} = \frac{\partial  \ln  y}{\partial  \ln \xi}$. Consequently, the second and last step is to solve for $ \frac{d y}{d \xi}$:
\begin{align}
\frac{d y}{d \xi} = \frac{(9x+10)y}{5\xi}. \label{ode1}
\end{align} 

To obtain the second ODE, thus arriving at an expression for $ \frac{d x}{d \xi}$,  an analogous transformation of Eq. \eqref{transpde}  has to be performed, which results in 
\begin{align}
\frac{d x}{d \xi}=-\frac{\left ( f-\xi \right)\left( \frac{9}{5}x+2\right)+\kappa \xi}{\xi f'} . \label{dx}
\end{align} 
Contrary to the case \citetalias{illenseer2015self} faced, the ODEs are not coupled, i.e. we can solve Eq. \eqref{dx} independently of Eq. \eqref{ode1}.

\subsection{Auxiliary conditions} \label{aux_con}
To arrive at a well-defined problem, it is necessary  to specify auxiliary conditions to solve the initial-boundary-value problem which the PDE \eqref{transevo} poses \citep[e.g.][]{ince}. The PDE describes the dynamical development of $\Omega(\varpi, \tau)$, i.e. the evolution of the angular velocity field of an accretion disk. Obviously, the spatial domain of physical relevance is given by $0< \varpi \leq \infty$. Consequently, two boundary values and one initial condition $\Omega_0(\varpi)$ at time $\tau=0$ are needed.  

Furthermore, these conditions have to be consistent with the properties of the physical system which is modeled to arrive at meaningful results. The model demands a ubiquitously positive surface density $\Sigma$ and enclosed mass $M$. In addition, $\partial_r M \geq 0$ is required.  Since radial momentum transport denoted by
\begin{align}
r^3\Omega^2=GM(r)>0 \: \: \: \rightarrow \: \: \: \Omega \not= 0 \label{aux_mass}
\end{align}   
holds for this model \citepalias{illenseer2015self}, the auxiliary conditions must also comply with 
\begin{align}\label{aux}
\frac{\partial \, \ln \, r^3\Omega^2}{\partial \, \ln \, r}=2x+3\geq0 \: \: \: \rightarrow \: \: \:  x \geq -\frac{3}{2}
\end{align} 
in the relevant domains of the parameters (ibid.).
\subsubsection{Initial conditions}
To arrive at an initial condition for $\Omega_0$, it is useful to take a look at Eq. \eqref{invariants}, which describes the group invariants. After solving for $\Omega$ one can take the limit and arrive at the following relation: 
\begin{align}
\Omega_0(\varpi)=\varpi^{(\frac{\kappa-2}{3})}\lim_{\xi \rightarrow \infty}(y^{\frac{1}{3}}\xi^{\frac{-\kappa}{3}}). 
\end{align} 
Since $\Omega_0$ should not vanish (Eq. \eqref{aux_mass}), it follows that
\begin{align}
y(\xi) \propto \xi^{\kappa} \: \: \:  \text{for} \: \: \:  \xi \rightarrow \infty,
\end{align} 
which consequently leads to the initial condition for self-similar solutions being a power law of radius
\begin{align} \label{omega0}
\Omega_0(r) &\propto r^{\frac{5}{3}(\frac{\kappa-2}{3})},
\end{align} 
where the second expression follows from the definition of $\varpi$ in Eq. \eqref{defdelta}. 

As \citetalias{illenseer2015self} point out, rotation laws of the kind of $\Omega \propto r^{\mu}$ cause infinite centrifugal forces for $r \rightarrow \infty$, if the exponent $\mu$ exceeds $-\frac{1}{2}$. Additionally, the monopole approximation used in the model breaks down when $\mu$ passes $-1$ and approaches $-\frac{1}{2}$. However, for up to $\mu = -\frac{3}{4}$ the error still appears acceptable (ibid.). After identifying $\mu$ with $\frac{5}{3}(\frac{\kappa-2}{3})$, this imposes an upper limit on $\kappa$. Equations \eqref{transx} and \eqref{aux} with \eqref{omega0} yield 
\begin{align}
-\frac{3}{2}\leq x \leq \frac{5}{3}\left(\frac{\kappa-2}{3}\right)
\end{align} 
as a lower limit for the exponent of the power law. Therefore,
\begin{align}
-\frac{3}{2}\leq\frac{5}{3}\left(\frac{\kappa-2}{3}\right) \leq -\frac{3}{4} \label{kappa_range}
\end{align} 
holds, which translates to a domain of $\kappa$ of 
\begin{align}
-\frac{7}{10} \leq \kappa \leq \frac{13}{20}.
\end{align} 
\subsection{Boundary conditions}
The dependence\footnote{Owing to the definition of $\varpi$ in Eq. \eqref{defdelta}, the behaviour of $\varpi$ corresponds to the behaviour of $r$.}  of $\xi$ on $\varpi$ and $\tau$ expressed in Eq. \eqref{invariants} yields the following:
\begin{align}
\xi \rightarrow 0 &\Leftrightarrow
 \begin{cases} \tau \rightarrow \infty   \: \: \:  &\text{for any fixed, positive} \, \varpi \\
 \varpi \rightarrow 0  \: \: \:  &\text{for any fixed, positive} \, \tau
\end{cases} 
\end{align} 
It is easy to see that the outer boundary condition coalesces with the initial condition. This reduction of auxiliary conditions is required by demanding self-similar solutions in terms of the independent variable $\xi$ \citep[][]{ames1965nonlinear}. 

At the inner rim of the disk, three reasonable boundary conditions exist: increasing, decreasing, and constant torque. The viscous torque is given by 
\begin{align}
\mathcal{G}(r,t)=2\pi r^2 \nu \Sigma r \frac{d\Omega}{dr},
\end{align} 
which, with the help of Eqs. \eqref{2}, \eqref{nu}, \eqref{sig}, \eqref{cs}, \eqref{gamma}, and \eqref{f} can be rewritten\footnote{The gravitational constant $G$ is set to unity.} and expressed in terms of the similarity variables to yield
\begin{align}
\mathcal{G}(r,t)&=\eta y \tau^{\kappa} f(x) \label{torque}.
\end{align}

The last step before being able to calculate physical meaningful solutions which correspond to specific physical situation is to specify a value for $\alpha$, $\kappa$, $\tau_0$, the central mass $M_{\star}$ at some time $\tau_0$, and torque $\mathcal{G}_{\star}$ at the inner rim at a specific time $\tau_0$. Expressing $M_{\star}$ and $\mathcal{G}_{\star}$ in terms of the group variants yields 
\begin{align}
M_{\star}(\tau)&=\tau^{\frac{2}{3}\kappa +\frac{7}{15}}\zeta_2^{\frac{2}{3}}, \; \; \; \; \; \;
\zeta_2= \frac{M_{\star \tau_0}^{\frac{3}{2}}}{{\tau_{0}^{\kappa +\frac{7}{10}}}},  \\
\mathcal{G}_{\star}(\tau)&=-\eta  \tau^{\kappa}\zeta_2\zeta_1, \; \; \; \; \; \;
\zeta_1=\frac{\mathcal{G}_{\star \tau_{0}}\tau_{0}^{\frac{7}{10}}}{-\eta M_{\star \tau_{0}}^{\frac{3}{2}}}
\end{align} 
with integration constants $\zeta_1$ and $\zeta_2$. To arrive at these results, an analysis of the phase plane and critical point is necessary. Since this analysis is, in principle, analogous to the analysis presented in \citetalias{illenseer2015self}, it was moved to Appendix A: \nameref{appendix}.

\section{Results}
All plotted solutions in this and the following section were obtained with the program \verb|lsode| \citep{lsode} using the BDF scheme. The value of $\xi_0$ can be chosen relatively arbitrarily, as long as the corresponding errors occurring during the calculation of the initial conditions via the linearized solutions do not exceed the inherent numerical errors. Before elaborating on the results in more detail, we note that $\xi$ is based on both  time and radius. Thus, each diagram can be read in two ways, i.e. for a fixed time the diagram shows the evolution of $x$ or $y$ according to radius, while for a fixed radius the diagram shows the time evolution of the respective dependent variable.

\subsection{Similarity solutions}
\begin{figure}[]
\resizebox{\hsize}{!}{\includegraphics{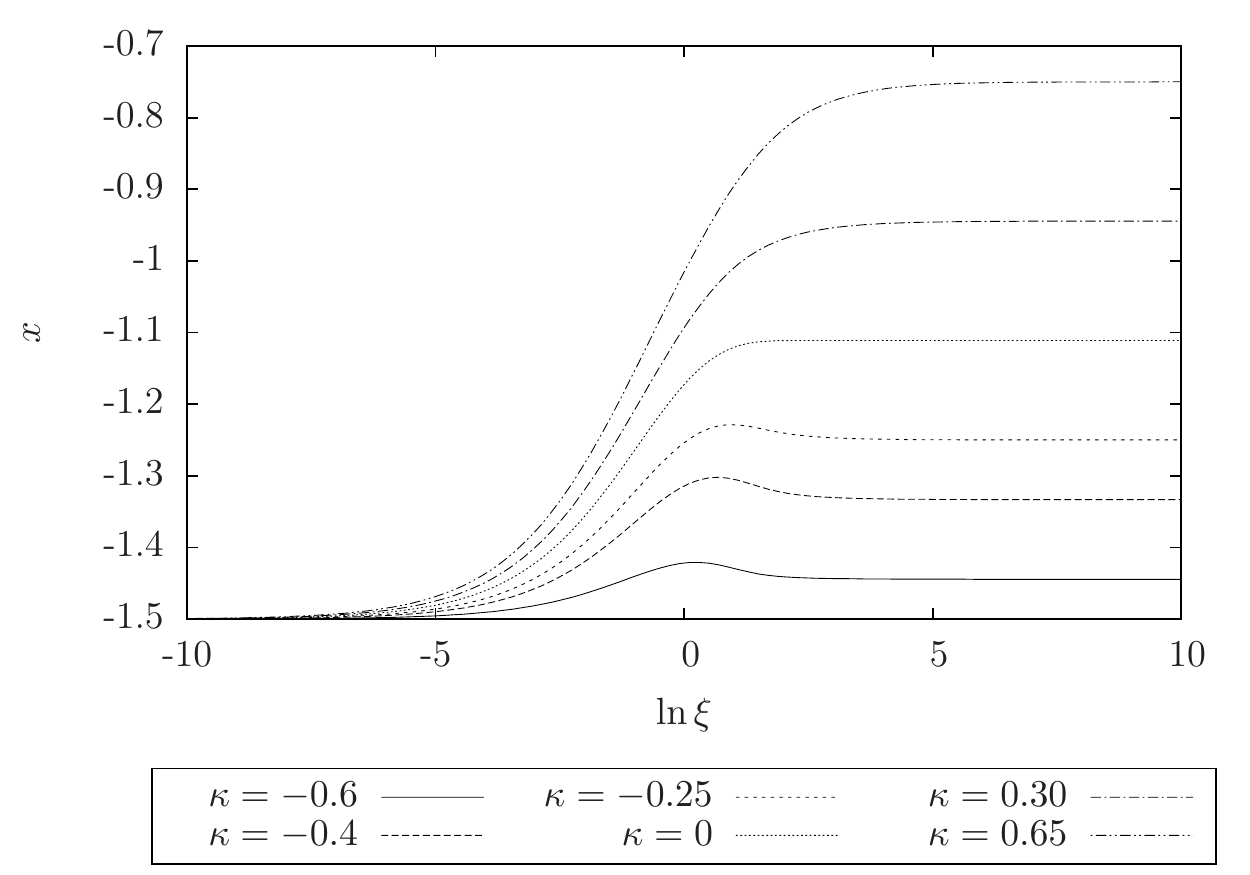}}
\caption{Solutions of $x$ for different values of $\kappa$.}
\label{x_kappa_var}
\end{figure}

\begin{figure}	
\resizebox{\hsize}{!}{\includegraphics{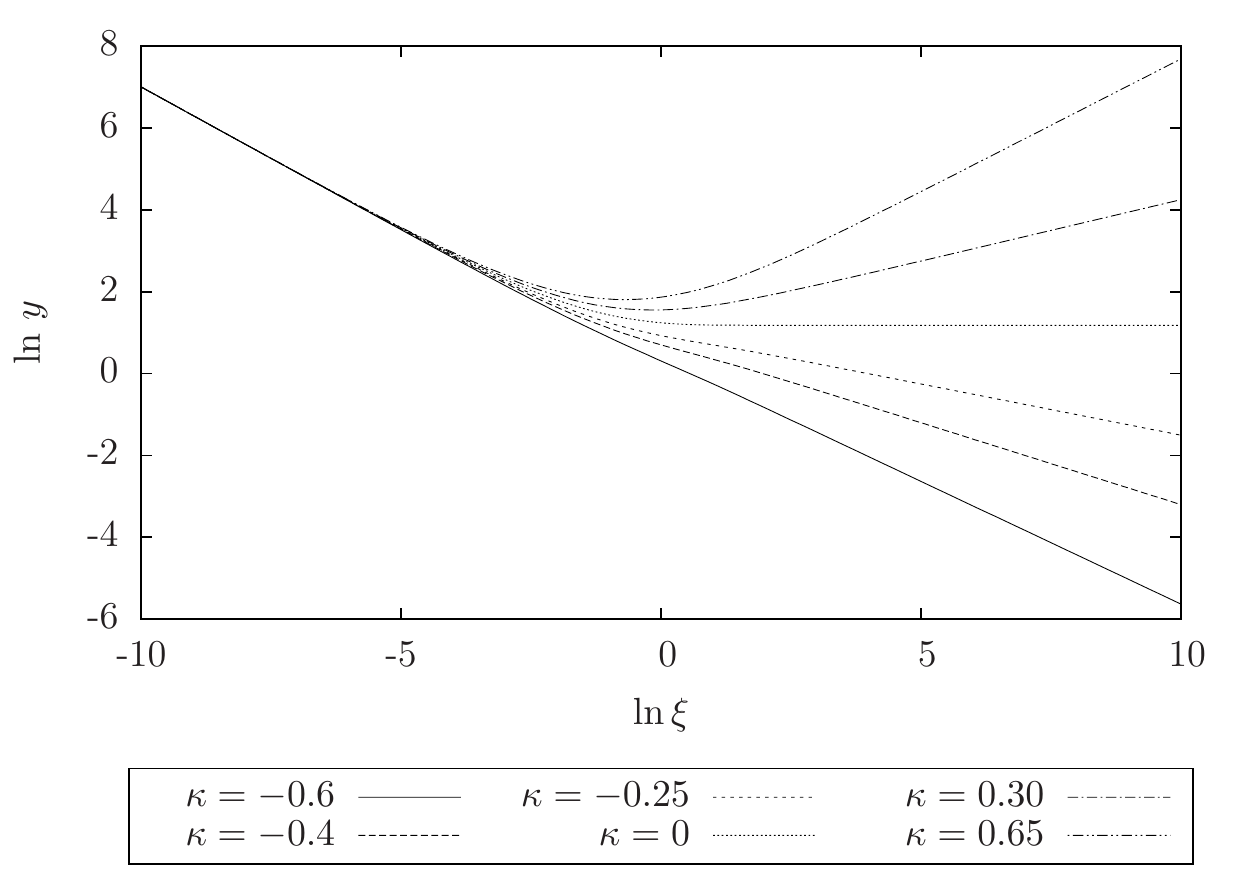}}
\caption{Solutions of $y$ for different values of $\kappa$.}
\label{y_kappa_var}
\end{figure}

\begin{figure}[]
\resizebox{\hsize}{!}{\includegraphics{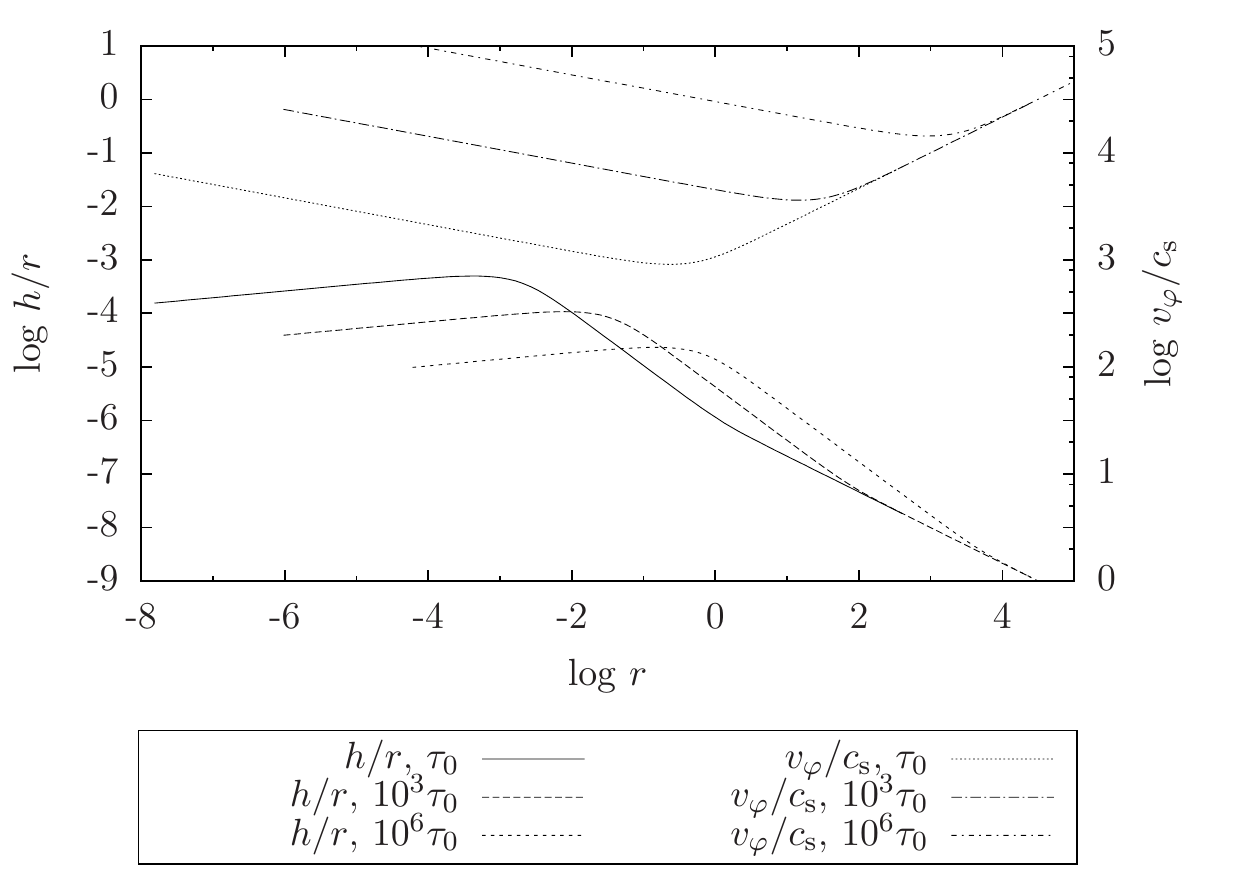}}
\caption{Solutions of $\frac{v_{\varphi}}{c_{\text{2}}}$ and $ \frac{h}{r}$ with $\alpha=0.01$ for $\kappa = 0.2$.}  \label{consis}
\end{figure}

Figures \ref{x_kappa_var} and \ref{y_kappa_var} show solutions for $x$ and $y$ (Eqs. \eqref{dx} and \eqref{ode1}) obtained with $\tau_0 = 1, M_\star(\tau_0) = 1, \ln\, \xi_0 = -8, \mathcal{G}_\star = 0$, and varying values of $\kappa$. 

The results depicted in Fig.  \ref{x_kappa_var} confirm the results from the last section and the Appendix. As predicted, $x$ approaches -1.5 for small values of $\xi$ for any value of $\kappa$ and grows to a somewhat higher finite limit for large values of  $\xi$ depending on the value of $\kappa$. The limit of $x$ for large $\xi$ depends on $\kappa$ (see Eq. \eqref{kappa_range}): 
\begin{align}
\lim_{\xi \rightarrow \infty}x=\frac{5}{3}\left(\frac{\kappa-2}{3}\right)  \label{x_limit}.
\end{align}
Using this result, it is possible to  find an approximation of $y$ for large radii:
\begin{align}
y&=\zeta_{2,\infty}\xi^{\kappa} \label{y(xi)_large_r}.
\end{align} 

Furthermore, the influence of the saddle point on the x-axis located at $x=-1.25$ is manifest in the form of a maximum for solutions with $\kappa \lesssim -0.25$, corresponding to the position of the saddle at $x=-1.25$. For solutions with $\kappa \geq -0.25$, the maximum becomes invisible because the limits of $x$ exceed -1.25. Consequently, these solutions do not approach the saddle point close enough to be influenced by it.  

A numerical solution for $\kappa = -0.7$ could not be computed owing to the function's stiffness in this regime. However, as  mentioned above, the solution for this case is a horizontal line at $x=-1.5$ corresponding to a fully evolved system with Keplerian rotation and no temporal evolution or radial dependence.

Taking a more physical perspective, the results are also satisfactory. Assuming constant time and a radial dependency of $x$, one arrives at the following picture: at the inner rim, self gravity can be neglected and rotation is consequently Keplerian. With growing radius the influence of self-gravity grows, leading to a flatter rotation law. Looking at the alternative picture of constant radius and time evolution the development of the disks becomes apparent. At a certain radius where rotation was initially governed by self-gravity, the rotation law shifts towards Keplerian rotation as a result of the mass transfer towards the inner parts of the disk through the accretion process.
In addition, Fig.  \ref{consis} shows that using the thin disk assumption and the slow accretion limit is justified because $v_{\varphi}$ is always highly supersonic, while the ratio $\frac{h}{r}$ is  $\ll 1$.

In addition, from the discussion above it becomes clear what $\kappa$ actually represents: the similarity parameter describes the mass distribution within the disk. 

Figure \ref{y_kappa_var} also confirms the results from the last section concerning Eq. \eqref{ode1}. For small values of $\xi$ all solutions are proportional to $\xi^{\frac{-7}{10}}$ while for large values of $\xi$ and no torque the value of $\kappa$ determines the slope of the linear term in Eq. \eqref{x(xi)}. Consequently, the value of $\kappa$ determines the value of $y$ for large $\xi$.

\subsection{Influence of the torque} \label{infl_t}

\begin{figure}[]
\resizebox{\hsize}{!}{\includegraphics{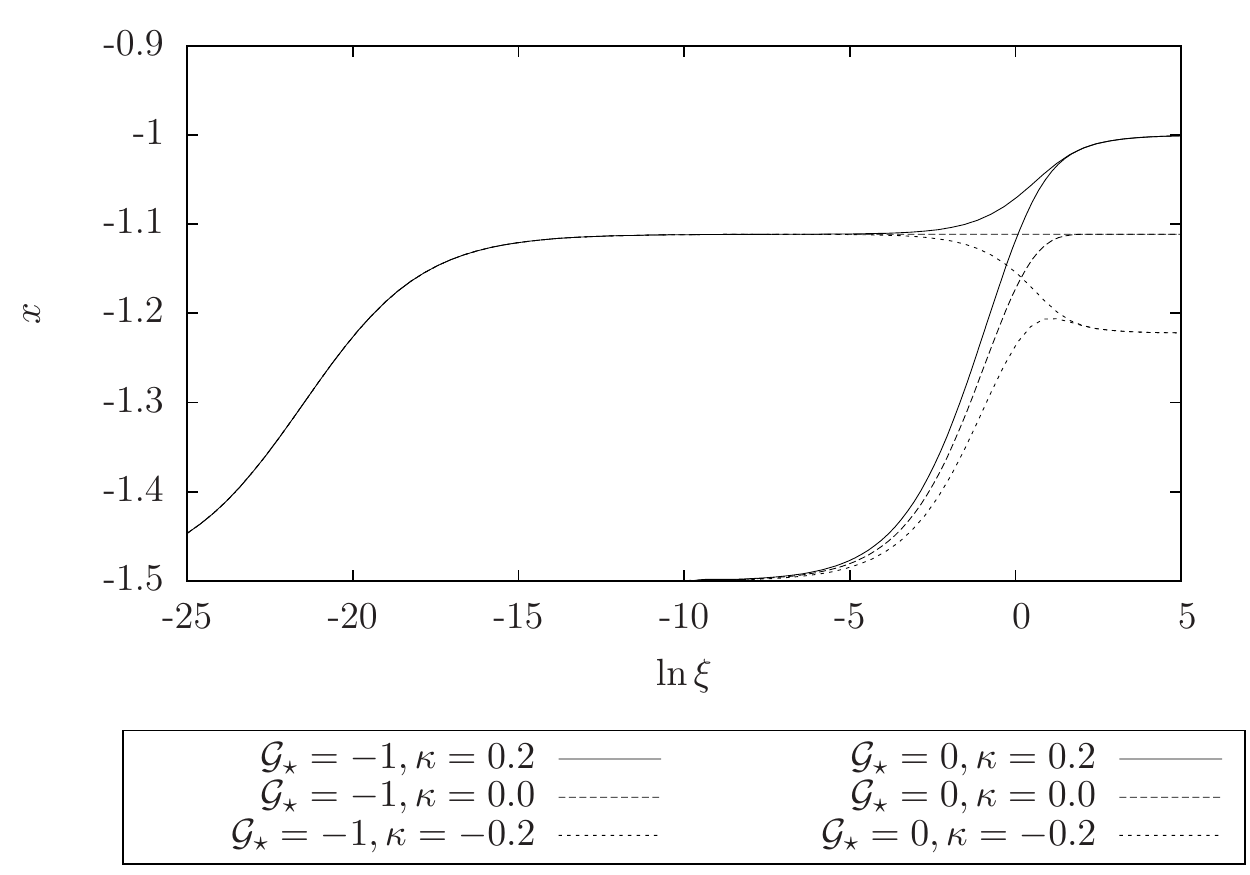}}
\caption{Solutions of $x$ for different torques $T_\star$ applied at the inner rim.}
\label{x_kappa_torque}
\end{figure}

All the solutions we have presented so far were obtained with no torque acting on the inner rim of the disk. In Fig.  \ref{x_kappa_torque}, the solutions were obtained using $\tau_0 = 1, M_\star(\tau_0) = 1, \ln\, \xi_0 = -25$, and the given $\mathcal{G}_\star$. 
However, the values in Figures \ref{dotm_r} and \ref{k_vari} depend explicitly on the value of $\eta$ and thus on the value of  $\alpha$, which was set to 0.01. The value was chosen since it lies well within the known range of $\alpha$ (0.1 to 0.001 \citep{king2007accretion}). 

A further mandatory ingredient is to calculate a numerical value for $\eta$. According to Eq. \eqref{gamma}, $\eta$ is dependent on the molecular weight. Assuming a gas composition of 75 \% helium and 25 \% hydrogen in a rather cold environment\footnote{In a cold environment  a neutral gas can be assumed.},  according to \cite{kippenhahn1990stellar} the mean molecular weight can be set to $ \approx 0.002 \frac{\text{kg}}{\text{mol}}$, which yields $\eta = \alpha^{\frac{4}{3}} \times 1.845 \times 10^{10} \,  \frac{\text{m}^{\frac{5}{3}}}{\text{s}}$. 

Using AGN scales  $\hat M = 10^3 \, M_{\odot}$, $\hat r = 1 \text{AU} $, and $\alpha = 0.01$,  a value of $1.5 \times 10^{-6}$ is obtained for $\eta$. This value is significantly lower than the value of $\beta = 10^{-3}$ used by \citetalias{illenseer2015self} and within the parameter range of $\alpha$ one can only arrive at $\eta = 3 \times 10^{-5}$. Consequently, one has to choose different scales to obtain $\eta$ within the range of $\beta$. Downsizing both scales until  a satisfactory value is reached leads to the scales of protoplanetary disks, while downsizing only one quantity leads either to a very  spread out disk or to a very small and massive disk. Both of the latter scenarios correspond to  Keplerian rotation and hence little evolution. To conclude,  as \cite{alpha_ineffective} point out, it seems that the $\alpha$ prescription is not able to describe the evolution of AGN faithfully. However, in Sect. \ref{appl_agn_disks} we   investigate further into the efficiency of the $\alpha$ prescription for AGN scales.

According to Eq. \eqref{torque}, the torque is increasing (positive $\kappa$), decreasing (negative $\kappa$), or constant ($\kappa$ = 0). In all three cases, the transition from the Keplerian to the self-gravitating regime starts at significantly smaller radii with respect to  earlier times than in the solution obtained with zero torque. A significant difference to the no-torque solution is that the solutions now have a third, intermediate step which serves as a maximum for the negative $\kappa$ and as a saddle for positive $\kappa$.

\subsection{Comparison of viscosity prescriptions}
\begin{figure}[]
\resizebox{\hsize}{!}{\includegraphics{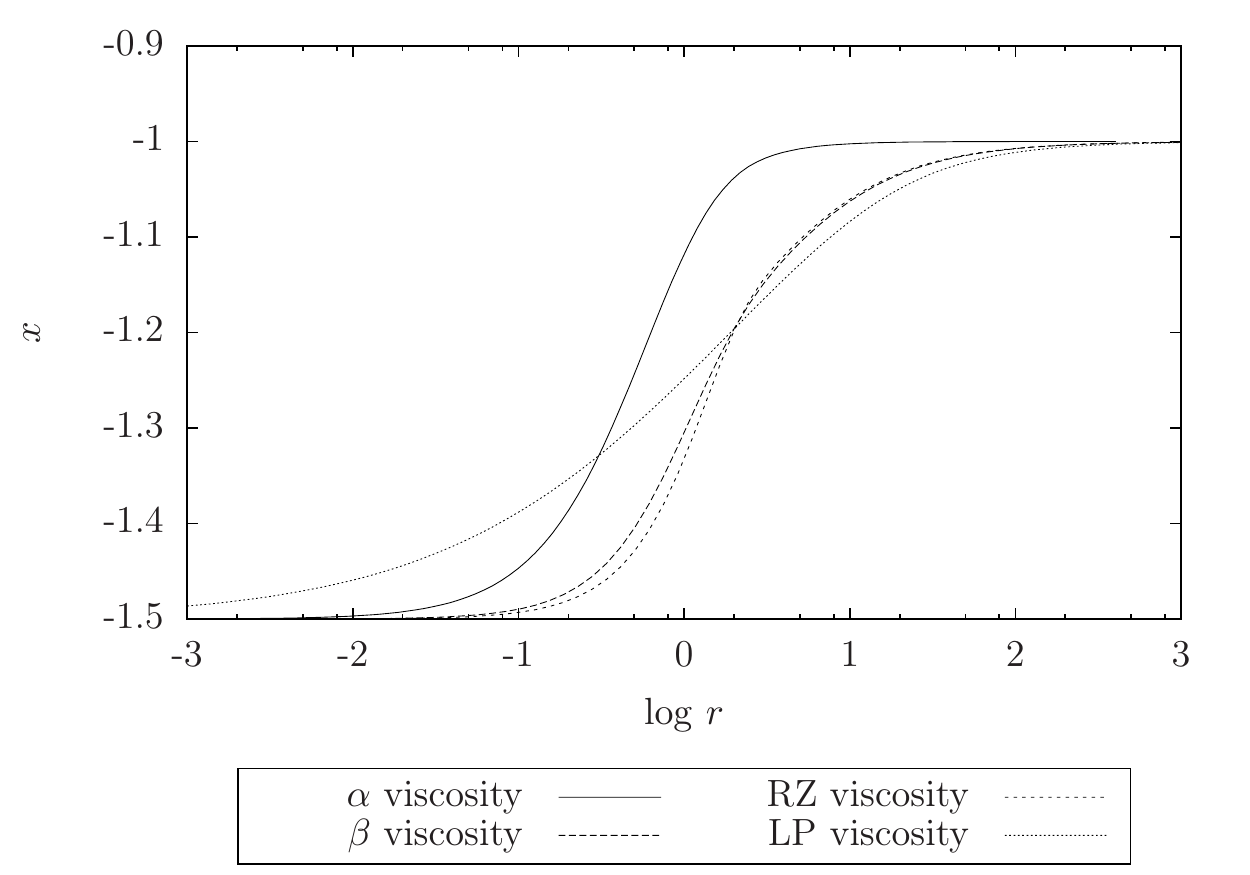}}
\caption{Solutions of $x$ for different viscosity prescriptions.}
\label{x_visc_var}
\end{figure}

To be able to compare our results with the results from \citetalias{illenseer2015self}, it is necessary to calculate $r$ from $\xi$ because the definitions of the similarity variable differ. Since the definitions of $y$ also differ substantially, it is only reasonable to compare the different dependencies of $x$ on the radial coordinate $r$. To avoid the influence of a viscosity parameter, only solutions  with zero torque are compared since they are still independent of a viscosity parameter  for all viscosity prescriptions. Thus, Fig.  \ref{x_visc_var} shows solutions for $x$ with initial conditions $\tau_0 = 1, M_\star(\tau_0) = 1, \ln\, \xi_0 = -8, \mathcal{G}_\star = 0$, and the same or the corresponding values\footnote{This is easily verified if one considers that all solutions have the same limit for large radii.} of the similarity parameter $\kappa$ for four different viscosity prescriptions. The relation between the $\kappa$ introduced in this paper and the $\kappa$ used by \citetalias{illenseer2015self} (hereafter $\tilde \kappa$) is $\frac{9}{5}\tilde \kappa + 2 = \kappa . \label{kappa_relation}$

Results for $\beta$-viscosity \citep[][(DSB)]{DSB}, RZ viscosity \citep{RZ}, and LP viscosity \citep{LP} were taken from \citetalias{illenseer2015self}. In general, the results are  similar, i.e. all solutions are basically step functions with an  abrupt transition marking the change from a Keplerian to a self-gravitating rotation regime. All in all, the $\alpha$ prescription seems to take a middle position between LP viscosity, to which it is most closely related, and $\beta$ with respect to RZ viscosity. 

\section{Discussion}

\begin{figure}[]
\resizebox{\hsize}{!}{\includegraphics{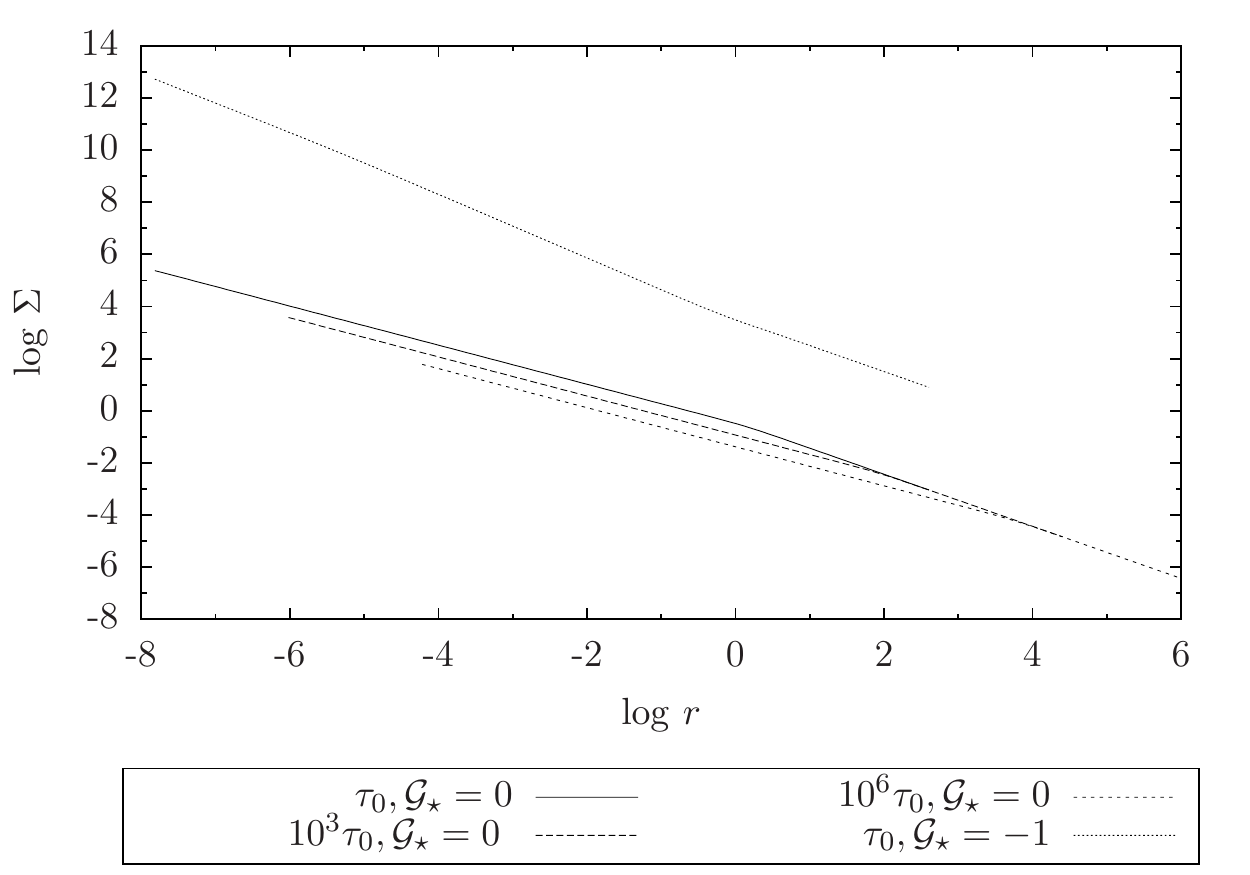}}
\caption{Solutions of $\Sigma(r)$ for different times and torques $\mathcal{G}_\star$.}  \label{sigma_r}
\end{figure}

\begin{figure}[]
\resizebox{\hsize}{!}{\includegraphics{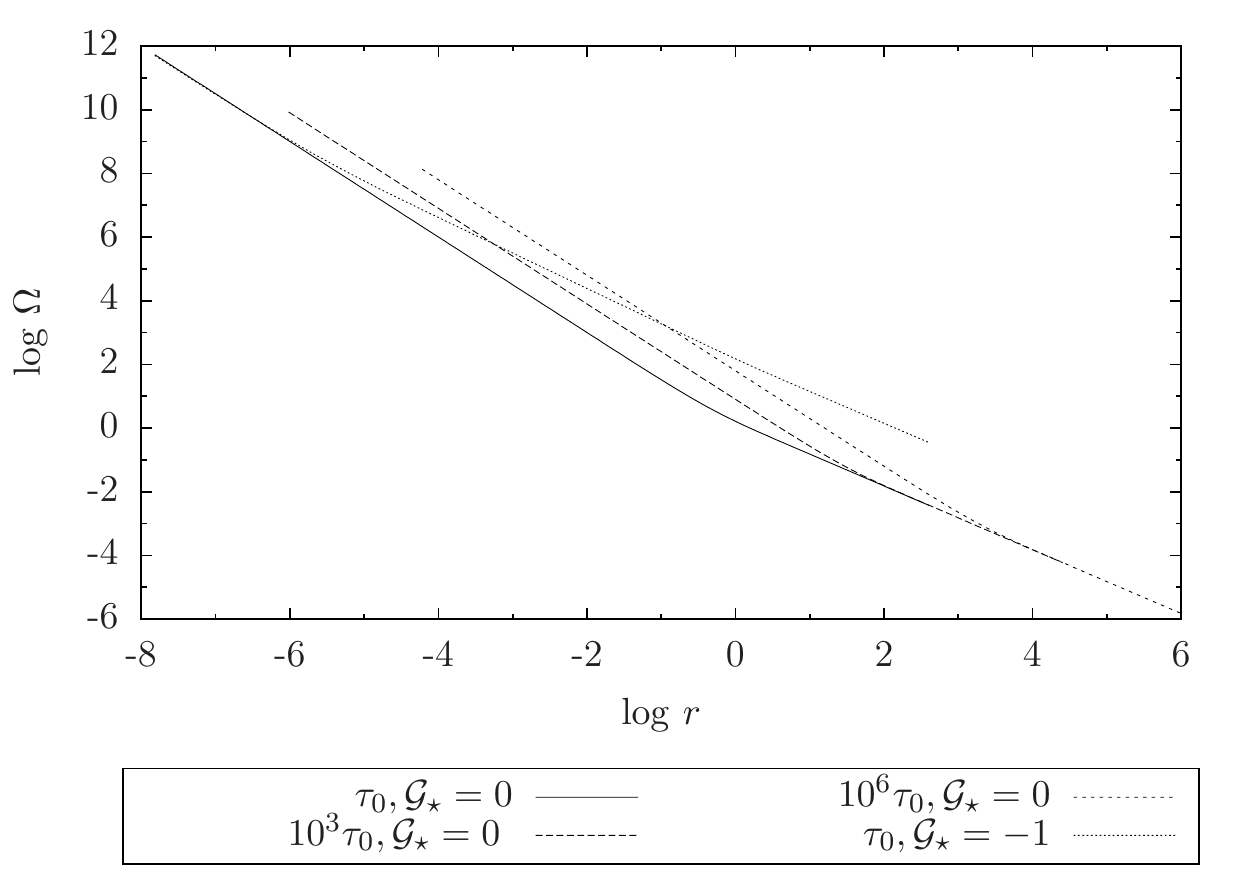}}
\caption{Solutions of $\Omega(r)$ for different times and torques $\mathcal{G}_\star$.}  \label{omega_r}
\end{figure}

\begin{figure}[]
\resizebox{\hsize}{!}{\includegraphics{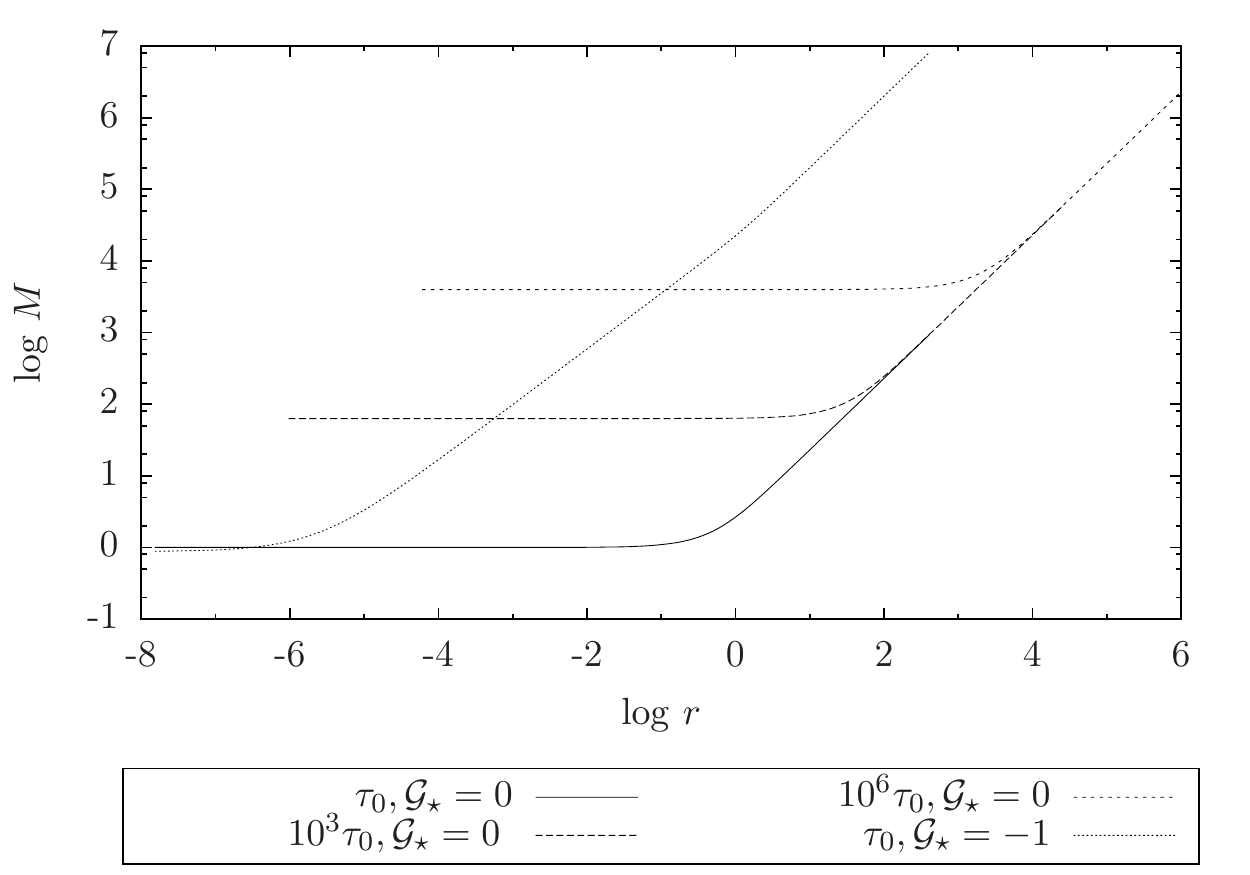}}
\caption{Solutions of $M(r)$ for different times and torques $\mathcal{G}_\star$.}  \label{m_r}
\end{figure}

\begin{figure}[]
\resizebox{\hsize}{!}{\includegraphics{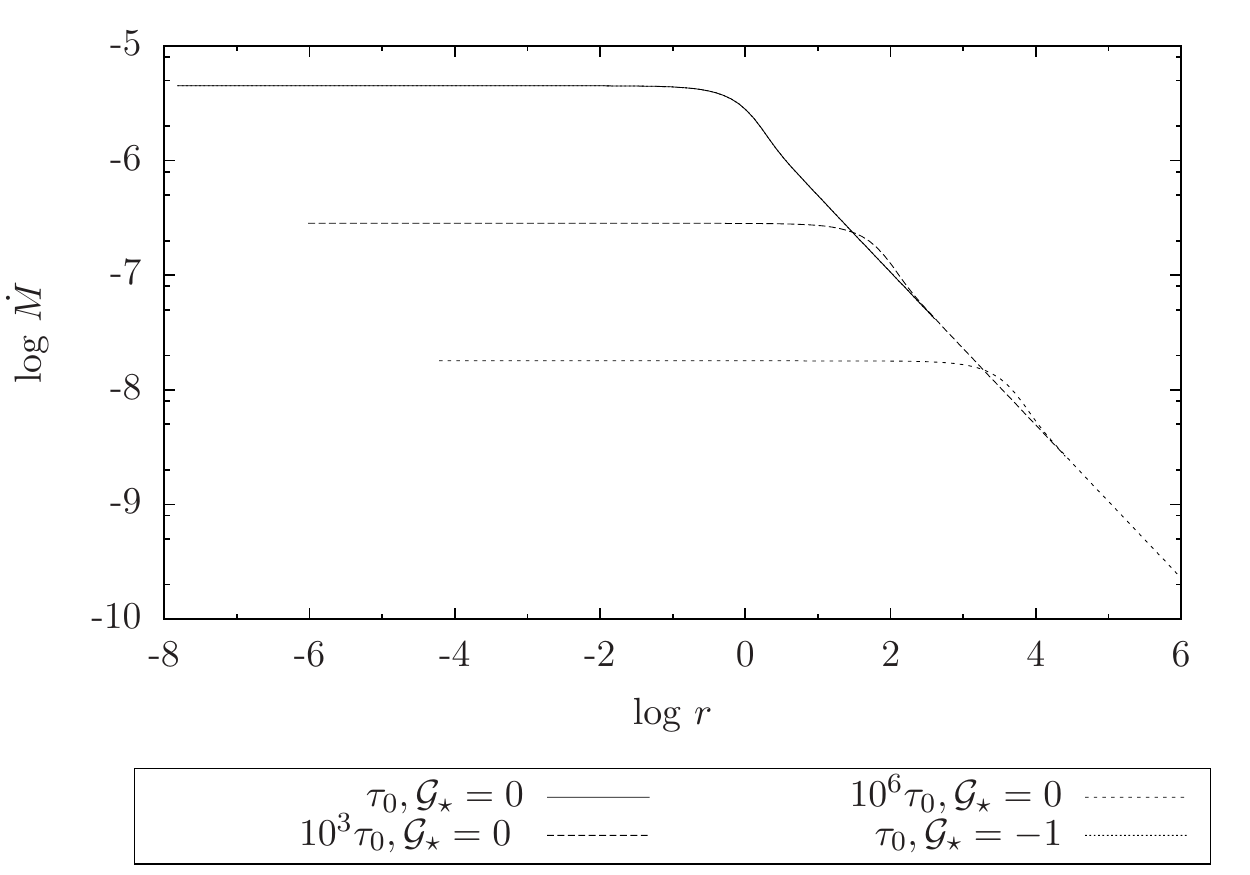}}
\caption{Solutions of $\dot M(r)$ with $\alpha=0.01$ for different times and torques $\mathcal{G}_\star$.}  \label{dotm_r}
\end{figure}

\begin{figure}[]
\resizebox{\hsize}{!}{\includegraphics{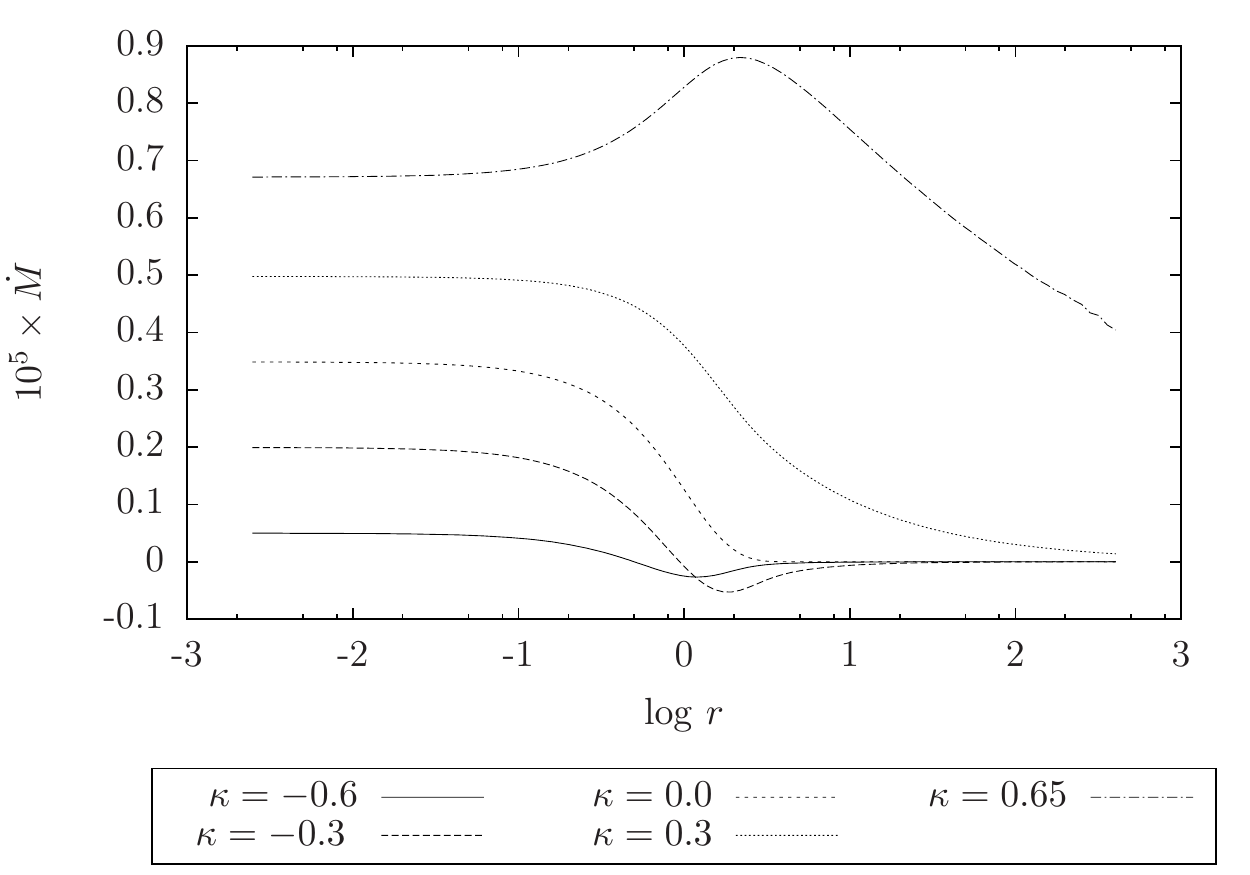}}
\caption{Solutions of $\dot M(r)$ with $\alpha=0.01$ for different values of $\kappa$.}  \label{k_vari}
\end{figure}

Using the definition of $\varpi$, $x$, and the invariants  in Eqs. \eqref{defdelta}, \eqref{transx}, and \eqref{invariants}, it is possible to express the defining physical quantities of the disk such as angular velocity $\Omega$, central mass $M$, and surface density $\Sigma$. Applying the solutions of $x$ and $y$,  the temporal and spatial development of these sizes  can now be investigated. The value of 
$\Omega$ can be directly derived via solving the definition of $y$ for $\Omega$ and substituting $\varpi$:
\begin{align}
\Omega = y^{\frac{1}{3}} \xi^{\frac{-2}{3}} \tau^{\frac{\kappa}{3}-\frac{2}{3}}. \label{Omega_invariants} 
\end{align} 
From the auxiliary conditions,  Eq. \eqref{aux_mass} can be used to obtain
\begin{align}
M = y^{\frac{2}{3}} \xi^{\frac{7}{15}} \tau^{\frac{2\kappa}{3}+\frac{7}{15}}. \label{M_invariants}
\end{align} 
In addition, the auxiliary conditions also yield the torque at the inner rim of the disk  expressed in terms of the similarity variables in Eq. \eqref{torque}.
For the surface density, it is simply necessary  to solve the relation $2 \pi G \Sigma = r \Omega^2 (2x+3)$ derived in \citetalias{illenseer2015self} for $\Sigma$ and insert the group invariants, which yields\footnote{The gravitational constant G is set to unity.}
\begin{align}
\Sigma = \frac{\tau^{\frac{2\kappa}{3}-\frac{11}{15}}  \xi^{\frac{-11}{15}} y^{\frac{2}{3}} (2x+3)}{2\pi} \label{Sigma_invariants}.
\end{align}

From these three basic quantities, the accretion rate $\dot M$ and radial velocity $v_{\text{r}}$ can be derived quite easily. Since $\frac{\partial \tau}{\partial t} = 5\eta$, the accretion rate is given by
\begin{align}
\dot M = 5 \eta \frac{M}{\tau} \left ( -\frac{6}{5}x - \frac{4}{3}+\frac{2}{3}\kappa \right ). \label{mdot_invariants}
\end{align} 

The same rationale was used to derive an expression for the radial velocity. Again,  starting with  a formula derived in \citetalias{illenseer2015self},   the  expressions derived above are inserted to arrive at the following relation:
\begin{align}
v_{\text{r}} = -\frac {5 \eta \xi^{\frac{3}{5}} \tau^{\frac{-2}{5}}  \left ( \frac{2\kappa}{3}-\frac{4}{3} - \frac{6x}{5} \right ) }{(2x+3)} \label{v_r_invariants}.
\end{align} 

All the necessary ingredients to find an expression for the vertically integrated viscous dissipation rate $Q_{\text{vis}}$ based on the similarity variables are now available and can be written as
\begin{align}
Q_{\text{vis}} = \frac{ \dot M M }{10 \pi \tau^{\frac{9}{5}}  \xi^{\frac{14}{5}}} \frac{(2x+3)^{\frac{4}{3}}  x^{\frac{4}{3}}}{ \left ( \frac{2\kappa}{3}-\frac{4}{3} - \frac{6x}{5} \right )}.
\end{align} 

It is a well-known and important result for Keplerian disks that in the radial direction $Q_{\text{vis}}$ is proportional to the product of central mass and accretion rate over the third power of radius and not explicitly dependent on the viscosity prescription \citepalias{shakura1973black}. To investigate whether this also holds for our self-similar model, we have to investigate the equation close to the critical point located at $x=-1.5$, which is equivalent to small values of $\xi$ and thus small radii. Applying the linearization of $x(\xi)$ for the no-torque condition in Eq. \eqref{x(xi)} to $(2x+3)^{\frac{4}{3}}$ and transforming the equation back to $r$ yields 
\begin{align}
Q_{\text{vis}} \approx \frac{ \dot M M }{10 \pi r^3} \frac{x^{\frac{4}{3}}}{ \left ( \frac{2\kappa}{3}-\frac{4}{3} - \frac{6x}{5} \right )}.
\end{align} 
Taking the limit for $x \rightarrow -\frac{3}{2}$ yields
\begin{align}
Q_{\text{vis}} &\approx \frac{  \dot M M }{10 \pi r^3} \lim_ {x \rightarrow -\frac{3}{2}} \frac{x^{\frac{4}{3}}}{ \left ( \frac{2\kappa}{3}-\frac{4}{3} - \frac{6x}{5} \right )} \notag \\
&= \frac{9}{4} \sqrt[3]{\frac{3}{2}} \frac{ 1 }{ \pi \left (10\kappa + 7 \right )} \frac{\dot M M }{r^3 } .
\end{align} 
This is a remarkable result since it confirms the validity of the model in the Keplerian regime. Furthermore, it is in concordance with the results presented in \citetalias{illenseer2015self}.

In the same fashion one can calculate the asymptotic behaviour of the other physical quantities for small and large radii and for initial conditions with and without torque supplied at the inner rim of the disk. The results of these calculations are shown in Table \ref{asymptotic_r}. For solutions containing $\kappa$, an equivalent solution with $\tilde \kappa$ is given for the sake of comparison. For $\Omega$ and $M$, the results are identical with the solutions \citetalias{illenseer2015self} obtained for DSB, RZ, and LP viscosity. The same is valid for $Q_{\text{vis}}$ and $\dot M$ in the limit $r \rightarrow 0$.  The results differ for the other quantities,  but generally speaking point in the same direction. However,  the radial velocity at the outer rim is independent of the similarity parameter, which poses an interesting deviation.   

\begin{table}
\centering
\begin{tabular}{ccccc}
\hline\hline   \\[-9pt] 
& \multicolumn{2}{c}{$r \rightarrow 0$} & \multicolumn{2}{c}{r $\rightarrow \infty$} \\
& zero torque & finite torque  &  \\
\hline   \\[-9pt] 
$\Omega$ & \multicolumn{2}{c}{$-\frac{3}{2}$} & $\frac{5\kappa}{9}-\frac{10}{9}$ & $\tilde \kappa$ \\[5pt]
$\Sigma$ & $-\frac{3}{4}$ & $-\frac{9}{8}$ & $\frac{10\kappa}{9}-\frac{11}{9}$& $2 \tilde \kappa + 1$ \\[5pt]
M & \multicolumn{2}{c}{0} & $\frac{10\kappa}{9}+\frac{7}{9}$ & $2 \tilde \kappa + 3$ \\[5pt]
$\dot M$ &  \multicolumn{2}{c}{0} & $\frac{10\kappa}{9}-\frac{8}{9}^{\dagger}$ & $2 \tilde \kappa + \frac{4}{3}^{\dagger}$ \\[5pt]
$v_{\text{r}}$ & $-\frac{1}{4}$ & $\frac{1}{8}$ & $-\frac{2}{3}^{\dagger}$ & $/$ \\[5pt]
$Q_{\text{vis}}$ & $-3$ & $-\frac{7}{2}$ & $\frac{20\kappa}{9}-\frac{26}{9}$& $4 \tilde \kappa + \frac{4}{3}$ \\[5pt]
$\mathcal{G}$ & $\frac{1}{2}$ & $0$ & $\frac{5\kappa}{3}$& $3 \tilde \kappa +\frac{10}{3}$ \\[5pt]
\hline \\[-9pt] 
\multicolumn{5}{c}{$\dagger$ for $\kappa = 0$ / $\tilde \kappa = -\frac{10}{9}$: exponential decay}
\end{tabular}
\caption{Power law exponents of the radial coordinate for small and large radii.} \label{asymptotic_r}
\end{table}

The solutions displayed in Fig.  \ref{sigma_r} and the following figures are given in non-dimensional units and are based on solutions obtained with $\tau_0 = 1, M_\star(\tau_0) = 1, \ln\, \xi_0 = -25$, and $\kappa = 0.2$. The value of $\kappa$ corresponds to $\tilde \kappa = -1$ which allows  the solutions to
be easily compared with those of \citetalias{illenseer2015self}. Solutions with a torque provided at the inner rim are dependent on $\alpha$ and were obtained with $\alpha = 0.01$ and $\eta =  1.5 \times 10^{-6}$, congruous with an AGN disk (see Sect. \ref{infl_t}). 

The results depicted in Fig. \ref{sigma_r} fit very well with the results noted in Table \ref{asymptotic_r}. All solutions have a kink where the transition between the Keplerian and the self-gravitating regime occurs which moves to regions further outward as time progresses, a behaviour already described by \cite{mineshige1996self} and \cite{mineshige1997self} for their self-similar $\alpha$ disk solutions. We note that the surface density  decreases through time and that the transition point  wanders to larger radii. This is a sensible result because the surface density has to decrease because the disk is losing mass due to accretion. Furthermore, since mass is moving inwards, the point where the mass of the disk becomes relevant for its gravitational potential moves outwards. In addition, the temporal development is consistent with the values given in Table \ref{asymptotic_t}. Moreover, it becomes apparent that a torque provided at the inner rim of the disk only affects this part of the disk. Such a torque allows a generally higher surface density but leads to a steeper decline in the Keplerian regime. The effect that the torque provides at the inner rim does not affect the proportionality in the self-gravitating regime, which is consistent throughout all quantities (Figures \ref{sigma_r} to \ref{dotm_r}), because  the approximation of $x$ for large radii (Eq. \eqref{x_limit}) does not contain a factor which differentiates between no-torque and finite torque. After a comparison of the figure  to the one given in \citetalias{illenseer2015self}, a qualitatively identical behaviour becomes evident. Moreover, the solutions in the Keplerian regime agree  with the well-known results for standard stationary $\alpha$ disks for the outer region \citepalias{shakura1973black}. This result is not surprising since the assumption that $T_{\text{eff}} = T_{\text{c}}$ made in Eq. \eqref{t_eff} is only true for the rather cold and thus optically thin outer parts of the disk.

The results concerning $\Omega$ are depicted in Fig.  \ref{omega_r}. Again, the transition point between regimes  moves outward with time for the same reason as in the previous case, but contrary to the case of $\Sigma$  -- consistently with the results from Table \ref{asymptotic_t} -- $\Omega$ grows with time. The value of $\Omega$  grows at a given radius because the central mass  grows through accretion. Thus, to compensate centrifugal forces, $\Omega$ has to grow. In addition, the results are also qualitatively and quantitatively consistent with those of \citetalias{illenseer2015self}. However, whether a torque is applied at the inner rim or not only seems to affect the position of the transition point.

The evolution of the contained mass visible in Fig.  \ref{m_r} is as expected from the literature \citep[e.g.][]{m_r}. The central mass  continually grows and the effect of a torque applied at the inner rim is that of an offset in the inner rim direction. Similar to the results obtained by \citetalias{illenseer2015self}, the case of the Kelperian valued similarity parameter belongs to a solution with constant disk mass. Although we cannot acquire numerical solutions for this case, the behaviour and dependencies are well known \citep[e.g. ][]{lynden1974evolution}.

While at first glance the results for accretion onto the central object depicted in Fig.  \ref{dotm_r} also seem to hold no surprises, they are in fact quite different from those obtained by \citetalias{illenseer2015self}. Depending on the similarity parameter, they report solutions with increasing, decreasing, and steady accretion rate. However, from Table \ref{asymptotic_t} and the range of $\kappa$ (Eq. \eqref{kappa_range}) it can be inferred that for $\alpha$-viscosity there are only  solutions with decreasing accretion rate $\dot M$, i.e. there are no quasi-stationary self-gravitating solutions. On the one hand,  this makes it easy to check whether the results -- when scaled to AGN dimensions -- exceed the Eddington limit, but it is also  questionable whether this model will be able to describe the full mass range \citep{fan2003survey} of SMBHs in the early universe \citep{duschl2011cosmogony}. 

However, our results for $\Omega$, $\Sigma$, and $M$ are consistent with the results of \cite{mineshige1996self}, i.e. we find the same radial power laws (see Table \ref{asymptotic_r}) if we set $\kappa = 0.2$. Moreover, from Table \ref{asymptotic_t} it can be  inferred that the temporal development of our accretion rate is consistent with \cite{mineshige1997self}. Furthermore, we find the same power laws for Keplerian rotation for $\Omega$ and $v_{\text{r}}$ in Table \ref{asymptotic_r} as \citetalias{shakura1973black} in their cold region, thus confirming that our more general model reproduces the results for standard stationary $\alpha$ disks.

\begin{table}
\centering
\begin{tabular}{ccccc}
\hline  \hline   \\[-9pt] 
& zero torque & finite torque & zero torque & finite torque   \\
\hline   \\[-9pt] 
$\Omega$ & \multicolumn{2}{c}{$\frac{\kappa}{3} + \frac{7}{30}$} &  \multicolumn{2}{c}{$\frac{3 \tilde \kappa}{5} + \frac{9}{10}$}  \\[5pt]
$\Sigma$ & $\frac{2\kappa}{3}-\frac{61}{60}$ & $\frac{2\kappa}{3}-\frac{19}{24}$ &$\frac{6 \tilde \kappa}{5} + \frac{19}{60}$ & $\frac{6 \tilde \kappa}{5} + \frac{13}{24}$ \\[5pt]
M &  \multicolumn{2}{c}{$\frac{2\kappa}{3}+\frac{7}{15}$} &  \multicolumn{2}{c}{$\frac{6 \tilde \kappa}{5} + \frac{9}{5}$} \\[5pt]
$\dot M$ &  \multicolumn{2}{c}{$\frac{2\kappa}{3}-\frac{8}{15}$} &  \multicolumn{2}{c}{$\frac{6 \tilde \kappa}{5} + \frac{4}{5}$} \\[5pt]
$v_{\text{r}}$ & $-\frac{1}{4}$ & $-\frac{19}{40} $ &\multicolumn{2}{c}{/} \\[5pt]
$Q_{\text{vis}}$ & $\frac{4\kappa}{3}-\frac{1}{15}$&$\frac{3}{10}$ &$\frac{12 \tilde \kappa}{5}+\frac{13}{5}$ &$/$  \\[5pt]
$\mathcal{G}$ & $\kappa - \frac{3}{10}$&$\kappa$ & $\frac{18 \tilde \kappa}{10} + \frac{17}{10}$& $\frac{9}{5}\tilde \kappa +2$ \\[5pt]
\hline  
\end{tabular}
\caption{Power law exponents of the time dependence for small radii.} \label{asymptotic_t}
\end{table}

The similarity parameter $\kappa$ has an influence on both, the temporal development of the accretion rate (Table \ref{asymptotic_t}) and, as one infers from Fig.  \ref{k_vari}, its total amount. Although, as Fig.  \ref{k_vari} indicates and Table \ref{asymptotic_r} predicts, the degree of self-gravity only affects the accretion process in the outer regions of the disk. 
Large values of $\kappa$, i.e. corresponding to self-gravity, yield the highest accretion rates and the slowest temporal decrease. Thus, the result of \citetalias{illenseer2015self} that ``objects embedded in self-gravitating disks with flatter rotation laws grow faster than those embedded in nearly Keplerian disks'' is confirmed. Moreover, our behaviour of $\dot M$ is qualitatively identical with the one reported for $\beta$-viscosity \citepalias{illenseer2015self};  depending on the value of the similarity parameter, $\dot M$ is either positive throughout the disk or -- at a certain radius -- becomes negative and converges towards zero, i.e. the self-gravitating part of the disk does not accrete material at all. For the $\alpha$ ansatz, the value of $\kappa$ separating the two regimes is $\kappa=0$ respectively $ \tilde \kappa = -\frac{10}{9}$, while for the $\beta$ ansatz it is $\tilde \kappa = -\frac{5}{4}$. Thus, $\alpha$ disks require an even more self-gravitating scenario than the $\beta$ ansatz to describe disks which accrete material at all radii.

\subsection*{Application to AGN disks} \label{appl_agn_disks}
Finally, we  return to physical dimensions to investigate whether the model complies with the scales known from observations.
Of the three scales involved in Eq. \eqref{dimless},  two can now be specified and  the third calculated. With these three basic scales  all the other scales (angular velocity, surface density, etc.)  can then be calculated. For an AGN example we choose $\hat M = 10^3 \, M_{\odot}$, $\hat r = 1 \text{AU} $, and $\alpha = 0.01$. The consequently arising scales are listed in Table \ref{agn_sizes}.

\begin{table}
\centering
\begin{tabular}{cccc}
\hline  \hline   \\[-9pt] 
$\hat r$ & $\hat M $ & $\hat t $ & $\tilde \eta$ \\
\hline   \\[-9pt] 
1 AU & $10^3 \, M_{\odot}$ & $5 \times 10^{-3}$ yr & $1.5 \times 10^{-6}$  \\
\hline  \hline   \\[-9pt] 
$\hat \tau $ & $\hat \Omega $ &  $\hat \Sigma$ & $\hat{\dot M} $ \\
\hline   \\[-9pt] 
1.189 s & $198.6 yr^{-1}$ &  $8.9 \times 10^{10} \frac{\text{kg}}{\text{m}^2}$ &  2 $\times 10^{5} M_{\odot} \text{yr}^-1 $ \\[5pt] 
\hline  
\end{tabular}
\caption{Scales for an AGN example with $\alpha = 0.01$.} \label{agn_sizes}
\end{table}

These scales can now be applied to the figures presented in this section. Applying\footnote{Through the definition of our scales in Eq. \eqref{dimless}, we also account for the numerical value of the gravitational constant $G$.} the scales to Figures \ref{dotm_r} and \ref{k_vari} shows that the accretion rate lies at $4.5 \times 10^{-6} \cdot 2 \times 10^{5} M_{\odot} \text{yr}^-1 = 0.9 M_{\odot} \text{yr}^-1$. For black holes $\lesssim 10^{7.7} M_{\odot}$ this value far exceeds the Eddington limit \citep{eddington1921} (assuming 10\% accretion effiency). However, for the late accretion phase of a relatively heavy black hole, this is a fitting result. 

Nevertheless, owing to the temporal decline of the accretion rate (Table \ref{asymptotic_t}) this is not sufficient to form SMBHs exceeding $10^9 M_{\odot}$ in less than $10^9 \text{yr}$ \citep{duschl2011cosmogony} -- even for the most self-gravitating systems -- as  can be seen from the following calculation:
\begin{align}
\int_{t_0=10^{7.7} \text{yr}}^{t_1=10^9\text{yr}} 0.9\, M_{\odot} \,(t \times 10^{-3}\text{yr})^{-0.1} dt &= 2.3 \times 10^8 M_{\odot} \\
2.3 \times 10^8 M_{\odot} + 10^{7.7} M_{\odot} &= 2.8 \times 10^8 M_{\odot} 
\end{align}  
Considering that the latest observations \citep{wu2015ultraluminous} suggest even more rapid growth of the mass of the central black hole, a lack of roughly an order of magnitude of mass is discouraging.

\section{Conclusions}
In addition to demonstrating the validity of the model in the Keplerian and in the self-gravitating case, the discussion in the previous section quite impressively points out the problem of the $\alpha$ prescription: it is too inefficient to faithfully describe the evolution of AGN containing SMBHs,  a result in agreement with \cite{alpha_ineffective}, \cite{mineshige1997self}, and \cite{Duschl:433740}. Since we  utilize a dynamic model, this result is not an artefact of the constraint that the accretion rate is constant. Furthermore, in contrast to \cite{Duschl:433740}, who have shown that $\alpha$-viscosity produces physically nonsensical results, we  developed a physically  consistent model which fails to produce the observed central masses within the time those observations suggest \citep{fan2003survey, wu2015ultraluminous}. Hence, our results weigh even more against the application of the $\alpha$ prescription to describe AGN. 

However, from a simple review of scales for which the $\alpha$ ansatz might produce sensible results, we find that those scales correspond to protoplanetary disks. As \cite{laughlin1994nonaxisymmetric} mention, $\alpha$-viscosity has already quite successfully been applied to model such disks, although mostly in steady models. Here, because  it is dynamic, our model poses an interesting alternative for future research. It is a possible vantage point for future research and development since  the admittedly rather simple treatment of the thermodynamic structure of the disk  would have to be modified. Furthermore,  the Eddington limit  could be incorporated directly into the model. 

Finally, all this yields one further conclusion. The viscosity prescription does not describe the actual viscous process, but parameterizes it, ultimately based on temperature. Obviously, our understanding of the physical processes governing the temperature within an AGN disk are too poor to give the correct relation. However, since that relation seems to be a better assumption for protoplanetary disks than for AGN disks, this hints at hitherto unknown or unconsidered physics in AGN. 

\begin{appendix}  
\section{Phase plane and critical point} \label{appendix}
\begin{figure}[] 
\centering
\resizebox{\hsize}{!}{\includegraphics{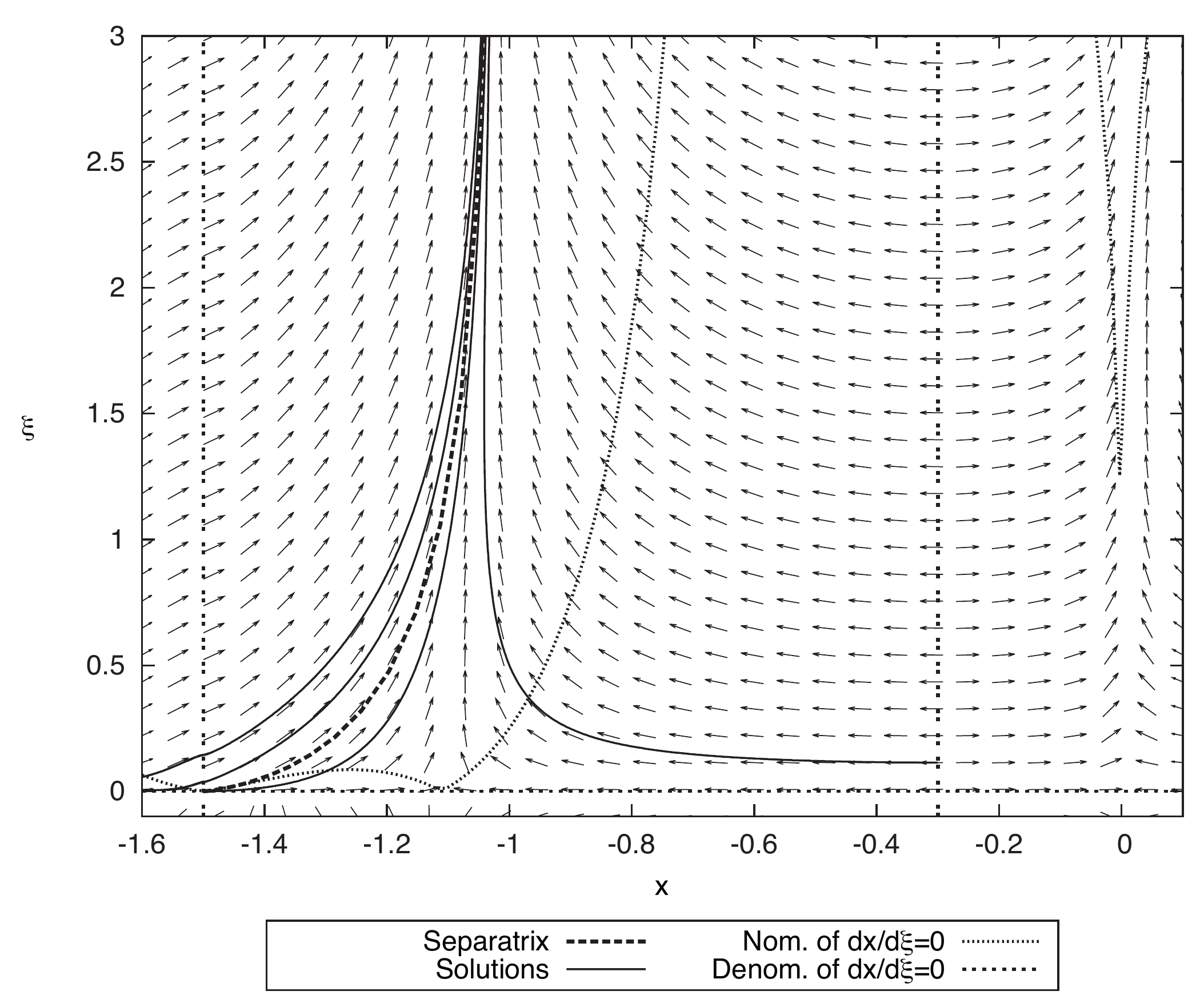}} 
\caption{Phaseplane of $\frac{d x}{d \xi}$ for $\kappa = 0.2$}  
\label{phaseplane}
\end{figure}

After having determined the auxiliary conditions in Sect. \ref{aux_con} and taking into consideration that Eq. \eqref{dx} is independent of $y$, it is now time to have a first look at the phase plane of the equation in Fig. \ref {phaseplane} to find out which points need further inquiry. Time and radius are always positive. Thus, according to our definition of $\xi$ (see Eq. \eqref{invariants}), $\xi$ is positive at any time. Since $x$ is defined as the local power law exponent of $\Omega$ (Eq. \eqref{x}), its range is limited to negative values and must be $\geq -1.5$ (the limit of Keplerian rotation) and $\leq -0.75$ (where the model breaks down). In general, the particulars of the phase plane  hinge on the value of $\kappa$. However, the general features remain the same and thus the value of $\kappa$ is set to $0.2$, which is well within the parameter range and is equivalent to the value of $\kappa = -1$ used by \citetalias{illenseer2015self} -- thereby often allowing convenient comparison.

The two dashed black vertical lines located at $x= -1.5$ and $ x=-0.3$ indicate the values of $x$ at which the function $f$ (Eq. \eqref{f}) becomes zero. Furthermore, the line at $x= -1.5$ shows the solution for $\kappa = -0.7$, which describes an already completely developed and thus completely Keplerian system. This is easily explained if one remembers that from the definition of  $\xi$ in Eq. \eqref{invariants} and the discussion of the auxiliary conditions one knows that a small $\xi$ corresponds to a small radius or late times. After the passing of sufficient viscous time scales, everything will have been accreted and consequently, lacking any self-gravity, a Keplerian rotation law is valid throughout the whole disk. In a similar fashion, the rotation law has to become  Keplerian  for small radii since the gravitational potential there is equivalent to a point mass potential.

Furthermore, the lines where the nominator respectively denominator of $\frac{dx}{d\xi}$ become zero and three critical points located on the x-axis at $x = -1.5, x = -1.25$, and $x=0$ are visible. Owing to the constraints of the model, only the critical points located at $x = -1.5, x = -1.25$ are potentially relevant for further analysis. 

In addition, Fig.  \ref {phaseplane} shows four numerically obtained (\verb|ode| \cite{ode}, Runge-Kutta scheme) solutions with varying initial condition and the line which probably defines the separatrix.  Since a small $\xi$ corresponds to a small radius, physically sensible solutions must approach $x=-1.5$ for small $\xi$ and, depending on the value of $\kappa$, a somewhat larger value of $x$ for large $\xi$ since the rotation law should reflect the growing self-gravity at the outer parts of the disk. From the four solutions depicted in Fig.  \ref {phaseplane} only one solution appears to meet these requirements, i.e. the solution which seemingly enters the critical point located at $(-1.5\mid 0)$. In order to be able to deliberately generate physically meaningful solutions, one has to be able to determine precise initial conditions. To find these conditions, the next step is to analyse the precise nature of the critical point. 

By the same token a more sophisticated analysis of the system around the critical point located at $x=-1.25$ can be neglected because a physically sensible solution will not be in its vicinity. Classifying  it as a saddle point \citep[e.g.][]{ince} is sufficient.
\section*{Critical point}
Although the phase plane presented above already shows satisfying results, one still has to determine the behaviour of the system in the vicinity of the critical point located at $(-1.5\mid 0)$ to be able to generate solutions corresponding to specific physical situations. Via standard methodology \citep[e.g.][]{ince}, the critical point can easily be classified qualitatively as an unstable node. Solutions which enter the critical point approach it from an area between the x-axis and the yet to be determined separatrix, i.e. the two characteristic directions \cite{frommer}. However, further quantitative results are necessary to arrive at sensible initial conditions. In order to achieve these conditions, one can linearize the problem in the vicinity of the critical point.

To begin with, let us consider Eq. \eqref{dx}. Owing to its highly non-linear dependence on variable $x$, which cannot be approximated, the equation is transformed to  
\begin{align}
\frac{d f}{d \xi}=-\frac{\left ( f-\xi \right)\left( \frac{9}{5}x(f)+2\right)+\kappa \xi}{\xi} . \label{df}
\end{align} 
and $x$ is now considered a function of $f$. Now, one can use a Taylor series to approximate $f^3(x)$ at $x=-1.5$, solve for $x$, and utilize $f=0$ at the critical point to arrive at
\begin{align}
x\approx-\frac{3}{2}+\frac{1}{2}\left( \frac{3}{2}\right)^{-\frac{1}{4}}(-f)^{\frac{3}{4}}.  \label{taylor}
\end{align}
The consequent simplification by neglecting terms of higher order in Eq. \eqref{df} leads to a formula which can actually be integrated analytically and solved for $f$ which gives 
\begin{align}
f=-\left (\frac{10}{3}\kappa +\frac{7}{3} \right )\xi+\zeta_1 \xi^{\frac{7}{10}} \; \; \; \; \; \; (\text{for\,} \xi \ll 1), 
\end{align} 
where $\zeta_1$ denotes the integration constant yet to be determined. To arrive at an equation including $x$, one simply has to use Eq. \eqref{taylor} to convert $f$ back to $x$:
\begin{align}
x=-\frac{3}{2}+\frac{1}{2}\left( \frac{3}{2}\right)^{-\frac{1}{4}}\left \{ \left (\frac{10}{3}\kappa +\frac{7}{3} \right )\xi-\zeta_1 \xi^{\frac{7}{10}} \right \}^{\frac{3}{4}} \; \; \; \; \; \; (\text{for\,} \xi \ll 1). \label{x(xi)}
\end{align} 
This solution shows that $\zeta_1 \xi^{\frac{7}{10}}$ is the dominant term close to the critical point where $\xi \rightarrow 0$ as long as $\zeta_1 \neq 0$. For $\zeta_1 = 0$, only the linear term remains and denotes the separatrix. All solutions for  $\zeta_1 \neq 0$ will eventually converge to the solution of the linearized equation. 

To solve Eq. \eqref{ode1} in the vicinity of the critical point,  one can drop the $\xi$ dependent term in Eq. \eqref{x(xi)}, and insert the result ($x=-1.5$)  into the aforementioned equation, which can now be easily integrated to obtain
\begin{align}
y&=\zeta_2\xi^{-\frac{7}{10}}. \label{y(xi)}
\end{align} 
What remains to be done is to define the integration constants $\zeta_1$ and $\zeta_2$. 

For $\zeta_2$, one can use that Eq. \eqref{aux_mass} gives a relation to the central mass for a certain point in time if $r \rightarrow 0$:
\begin{align}
M_{\star}(\tau)=\lim_{r \rightarrow 0}r^3\Omega^2
\end{align} 
If we express this with the help of the group invariants in Eq. \eqref{invariants} and the definition of $y$ in Eq. \eqref{y(xi)}, we obtain
\begin{align}
M_{\star}(\tau)=\tau^{\frac{2}{3}\kappa +\frac{7}{15}}\zeta_2^{\frac{2}{3}}, \; \; \; \; \; \;
\zeta_2= \frac{M_{\star \tau_0}^{\frac{3}{2}}}{{\tau_{0}^{\kappa +\frac{7}{10}}}} . \label{ceta2}
\end{align} 
Thus, if one specifies $\kappa$ and the central mass $M_{\star \tau_0}$ at some time $\tau_0$ one can calculate $\zeta_2$.

To compute the second integration constant $\zeta_1$, the viscous torque at the inner boundary of the disk will be of interest. There are two sensible possibilities for any time $0<t<\infty$, i.e. vanishing and finite torque. To analyze the limit at the inner rim, it is helpful to express $y$ in Eq. \eqref{torque} via $x$. In consequence, one has to eliminate $y$ and subsequently $\xi$. The first step is easily done using Eq. \eqref{y(xi)}:
\begin{align}
\mathcal{G}&= \eta  \tau^{\kappa}\zeta_2\xi^{-\frac{7}{10}} f(x).  \label{G_trans}
\end{align} 
For the second step, one has to solve Eq. \eqref{x(xi)} for $\xi$ with $\xi \ll 1$, which gives two solutions; one for $\zeta_1 = 0$ and one for $\zeta_1 \neq 0$:
\begin{align}
\zeta_1 \neq 0 &\; \;  \rightarrow  \; \; \xi = \left ( \frac{3}{2}\right )^{\frac{10}{21}} \frac{\left (2x+3 \right )^{\frac{40}{21}}}{\zeta_1^{\frac{10}{7}}}  \label{xi_x_0}
\\
\zeta_1 = 0 &\; \; \rightarrow \; \;  \xi = \left ( \frac{3}{2}\right )^{\frac{1}{3}} \frac{\left (2x+3 \right )^{\frac{4}{3}}}{\left (\frac{10}{3}\kappa +\frac{7}{3}\right )}   \label{xi_x_1}
\end{align} 
From Eq. \eqref{G_trans}, one can infer that a vanishing torque at the inner boundary demands that 
\begin{align}
\lim_{\xi \rightarrow 0} \xi^{-\frac{7}{10}} f(x) = \lim_{x \rightarrow -\frac{3}{2}} \xi^{-\frac{7}{10}} f(x) = 0 .
\end{align} 
If one plugs in the solution from \eqref{xi_x_1} and $f$ as defined in Eq. \eqref{f}, one can see that this is indeed the case:
\begin{align}
&\lim_{x \rightarrow -\frac{3}{2}} \left (\left ( \frac{3}{2}\right )^{\frac{1}{3}} \frac{\left (2x+3 \right )^{\frac{4}{3}}}{\left (\frac{10}{3}\kappa +\frac{7}{3}\right )}\right )^{-\frac{7}{10}} x^{\frac{1}{3}} (2x+3)^{\frac{4}{3}}  \notag \\
&=\left ( \frac{3}{2}\right )^{-\frac{7}{30}} \left (\frac{10}{3}\kappa +\frac{7}{3}\right )^{\frac{7}{10}} \lim_{x \rightarrow -\frac{3}{2}} x^{\frac{1}{3}} (2x+3)^{\frac{2}{5}} = 0.
\end{align} 
Hence, one can conclude that $\zeta_1 = 0$ denotes a special no-torque solution and defines the separatrix.

\begin{figure}
\resizebox{\hsize}{!}{\includegraphics{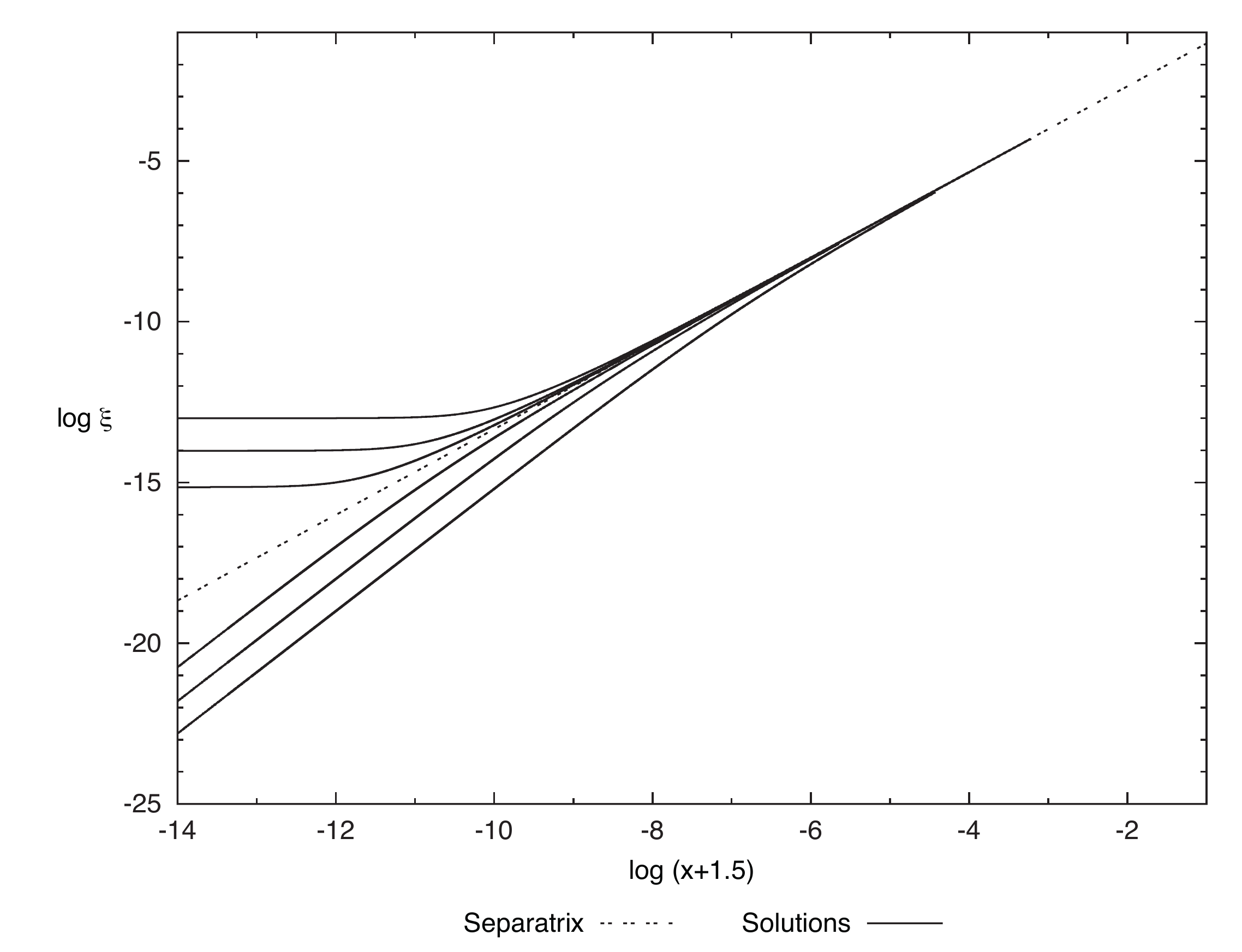}}
\caption{Logarithmic phaseplane of $\frac{d x}{d \xi}$ for $\kappa = 0.2$ }
\label{phaseplaneloglog}
\end{figure}

The behaviour of $\xi$ in the proximity of the critical point derived in this section can indeed be seen in Fig.  \ref{phaseplaneloglog}, where six solutions of Eq. \eqref{dx} and the separatrix are depicted. Three solutions were obtained using \verb|ode| \citep[][]{ode} with initial conditions above the separatrix and three solutions with initial conditions below the separatrix. The slope of the solutions below the separatrix is in concordance with the slope predicted in Eq. \eqref{xi_x_0} and the slope of the separatrix corresponds with the one given by Eq. \eqref{xi_x_1}, thereby confirming the results acquired with the linearized version of Eq. \eqref{dx}. 

Additionally, one can conclude that exactly one solution, i.e. the no-torque solution, enters the critical point along the singular characteristic direction defined by the separatrix and that infinitely many solutions enter the critical point along the other multiple characteristic direction, i.e. the x-axis \citep{frommer}. The initial conditions to obtain those solutions which enter the critical point along the x-axis are determined by the torque acting on the inner rim of the disk (Eq. \eqref{xi_x_0}). To obtain these conditions, one has to investigate the limit of $\mathcal{G}$ for $\xi \rightarrow 0$ and $\zeta_1 \neq 0$:
\begin{align}
&\lim_{x \rightarrow -\frac{3}{2}} \left (\left ( \frac{3}{2}\right )^{\frac{10}{21}} \frac{\left (2x+3 \right )^{\frac{40}{21}}}{\zeta_1^{\frac{10}{7}}}\right ) ^{-\frac{7}{10}} x^{\frac{1}{3}} (2x+3)^{\frac{4}{3}} \notag \\ 
&=\left ( \frac{3}{2}\right )^{-\frac{1}{3}} \zeta_1 \lim_{x \rightarrow -\frac{3}{2}} \frac{x^{\frac{1}{3}} (2x+3)^{\frac{4}{3}}}{\left (2x+3 \right )^{\frac{4}{3}}} = - \zeta_1.
\end{align} 
Thus, using Eq. \eqref{ceta2}, $\zeta_1$ can be calculated if one prescribes a torque $\mathcal{G}_{\star}$ at the inner rim at a specific time $\tau_0$ via 
\begin{align}
\mathcal{G}_{\star}(\tau)=-\eta  \tau^{\kappa}\zeta_2\zeta_1, \; \; \; \; \; \;
\zeta_1=\frac{\mathcal{G}_{\star \tau_{0}}\tau_{0}^{\frac{7}{10}}}{-\eta M_{\star \tau_{0}}^{\frac{3}{2}}}.
\end{align} 

\end{appendix}
\bibliographystyle{aa} 
\bibliography{masterthesis.bib} 

\begin{thebibliography}{35}
\expandafter\ifx\csname natexlab\endcsname\relax\def\natexlab#1{#1}\fi

\bibitem[{Ames(1965)}]{ames1965nonlinear}
Ames, W.~F. 1965, Nonlinear Differential Equations in Engineering (New York:
  Academic)

\bibitem[{{Balbus} \& {Hawley}(1991)}]{mri}
{Balbus}, S.~A. \& {Hawley}, J.~F. 1991, \apj, 376, 214

\bibitem[{Bluman {et~al.}(2010)Bluman, Cheviakov, \&
  Anco}]{bluman2010applications}
Bluman, G.~W., Cheviakov, A.~F., \& Anco, S.~C. 2010, Applications of symmetry
  methods to partial differential equations (Springer)

\bibitem[{Dresner(1998)}]{dresner1998applications}
Dresner, L. 1998, Applications of Lie's theory of ordinary and partial
  differential equations (CRC Press)

\bibitem[{Duschl \& Strittmatter(2011)}]{duschl2011cosmogony}
Duschl, W.~J. \& Strittmatter, P.~A. 2011, Monthly Notices of the Royal
  Astronomical Society, 413, 1495

\bibitem[{Duschl {et~al.}(1998)Duschl, Strittmatter, \& Biermann}]{DSB}
Duschl, W.~J., Strittmatter, P.~A., \& Biermann, P.~L. 1998, in Bulletin of the
  American Astronomical Society, Vol.~30, 917

\bibitem[{Duschl {et~al.}(2000)Duschl, Strittmatter, \&
  Biermann}]{Duschl:433740}
Duschl, W.~J., Strittmatter, P.~A., \& Biermann, P.~L. 2000, Astron.
  Astrophys., 357, 1123

\bibitem[{Eddington(1921)}]{eddington1921}
Eddington, A.~S. 1921, Zeitschrift f{\"u}r Physik A Hadrons and Nuclei, 7, 351

\bibitem[{Fan {et~al.}(2003)Fan, Strauss, Schneider, Becker, White, Haiman,
  Gregg, Pentericci, Grebel, Narayanan, {et~al.}}]{fan2003survey}
Fan, X., Strauss, M.~A., Schneider, D.~P., {et~al.} 2003, The Astronomical
  Journal, 125, 1649

\bibitem[{Frank {et~al.}(2002)Frank, King, \& Raine}]{frank2002accretion}
Frank, J., King, A., \& Raine, D. 2002, Accretion power in astrophysics
  (Cambridge University Press)

\bibitem[{Frommer(1928)}]{frommer}
Frommer, M. 1928, Die Integralkurven einer gew{\"o}hnlichen
  Differentialgleichung erster Ordnung in der Umgebung rationaler
  Unbestimmtheitsstellen (Springer)

\bibitem[{Goldreich \& Schubert(1967)}]{goldreich1967differential}
Goldreich, P. \& Schubert, G. 1967, The Astrophysical Journal, 150, 571

\bibitem[{Hindmarsh(1983)}]{lsode}
Hindmarsh, A.~C. 1983, IMACS transactions on scientific computation, 1, 55

\bibitem[{Illenseer \& Duschl(2015)}]{illenseer2015self}
Illenseer, T.~F. \& Duschl, W.~J. 2015, Monthly Notices of the Royal
  Astronomical Society, 450, 691

\bibitem[{Ince(1956)}]{ince}
Ince, E.~L. 1956, Ordinary differential equations. (Dover, New York)

\bibitem[{Kato {et~al.}(2008)Kato, Fukue, \& Mineshige}]{kato2008black}
Kato, S., Fukue, J., \& Mineshige, S. 2008, Black-Hole Accretion Disks --
  Towards a New Paradigm, Vol.~1

\bibitem[{King {et~al.}(2007)King, Pringle, \& Livio}]{king2007accretion}
King, A., Pringle, J., \& Livio, M. 2007, Monthly Notices of the Royal
  Astronomical Society, 376, 1740

\bibitem[{Kippenhahn {et~al.}(2012)Kippenhahn, Weigert, \&
  Weiss}]{kippenhahn1990stellar}
Kippenhahn, R., Weigert, A., \& Weiss, A. 2012, Stellar structure and evolution
  (Springer)

\bibitem[{L.~Filipov(1988)}]{m_r}
L.~Filipov, N.I.~Shakura, Y.~L. 1988, Advances in Space Research, 8, 163

\bibitem[{Laughlin \& Bodenheimer(1994)}]{laughlin1994nonaxisymmetric}
Laughlin, G. \& Bodenheimer, P. 1994, The Astrophysical Journal, 436, 335

\bibitem[{Lin \& Pringle(1987)}]{LP}
Lin, D. N.~C. \& Pringle, J.~E. 1987, Monthly Notices of the Royal Astronomical
  Society, 225, 607

\bibitem[{L{\"u}st(1952)}]{lust1952entwicklung}
L{\"u}st, R. 1952, Zeitschrift Naturforschung Teil A, 7, 87

\bibitem[{Lynden-Bell(1969)}]{bell}
Lynden-Bell, D. 1969, Nature, 223, 690

\bibitem[{Lynden-Bell \& Pringle(1974)}]{lynden1974evolution}
Lynden-Bell, D. \& Pringle, J.~E. 1974, Monthly Notices of the Royal
  Astronomical Society, 168, 603

\bibitem[{Mineshige \& Umemura(1996)}]{mineshige1996self}
Mineshige, S. \& Umemura, M. 1996, The Astrophysical Journal Letters, 469, L49

\bibitem[{Mineshige \& Umemura(1997)}]{mineshige1997self}
Mineshige, S. \& Umemura, M. 1997, The Astrophysical Journal, 480, 167

\bibitem[{{Richard} \& {Zahn}(1999)}]{RZ}
{Richard}, D. \& {Zahn}, J.-P. 1999, Astronomy and Astrophysics, 347, 734

\bibitem[{Salpeter(1964)}]{salpeter1964accretion}
Salpeter, E.~E. 1964, The Astrophysical Journal, 140, 796

\bibitem[{Shakura \& Sunyaev(1976)}]{shakura1976theory}
Shakura, N. \& Sunyaev, R. 1976, Monthly Notices of the Royal Astronomical
  Society, 175, 613

\bibitem[{Shakura \& Sunyaev(1973)}]{shakura1973black}
Shakura, N.~I. \& Sunyaev, R.~A. 1973, Astronomy and Astrophysics, 24, 337

\bibitem[{{Shlosman} {et~al.}(1990){Shlosman}, {Begelman}, \&
  {Frank}}]{alpha_ineffective}
{Shlosman}, I., {Begelman}, M.~C., \& {Frank}, J. 1990, \nat, 345, 679

\bibitem[{Tufillaro {et~al.}(1992)Tufillaro, Abbott, \& Reilly}]{ode}
Tufillaro, N.~B., Abbott, T., \& Reilly, J. 1992, An experimental approach to
  nonlinear dynamics and chaos (Addison-Wesley Redwood City, CA)

\bibitem[{Weizs{\"a}cker(1948)}]{weizsacker1948rotation}
Weizs{\"a}cker, C.~F. 1948, Zeitschrift f{\"u}r Naturforschung A, 3, 524

\bibitem[{Wu {et~al.}(2015)Wu, Wang, Fan, Yi, Zuo, Bian, Jiang, McGreer, Wang,
  Yang, {et~al.}}]{wu2015ultraluminous}
Wu, X.-B., Wang, F., Fan, X., {et~al.} 2015, Nature, 518, 512

\bibitem[{Zeldovich(1964)}]{zeldovich1964fate}
Zeldovich, Y.~B. 1964, Doklady Akademii Nauk SSSR, 155, 67

\end{thebibliography}

\end{document}